\documentclass[aps,prb,twocolumn,eqsecnum,amsmath,amssymb,floatfix,superscriptaddress]{revtex4}
\usepackage{tabularx}
\usepackage{bm}
\usepackage{subfigure}
\usepackage{euscript}
\usepackage{epsfig,psfrag,subfigure}
\usepackage{graphicx,psfrag,subfigure}
\usepackage{color}
\begin{document}
\title{Measuring fractional charge and statistics in fractional quantum Hall fluids through noise experiments}
\author{Eun-Ah Kim}
\affiliation{Department of Physics, University of Illinois at
Urbana-Champaign, 1110
W.\ Green St.\ , Urbana, IL  61801-3080, USA}
\affiliation{Stanford Institute for Theoretical Physics and Department of Physics, Stanford University,
 Stanford, CA 94305, USA}
 
\author{ Michael J. Lawler}
\author{ Smitha Vishveshwara}
\author{Eduardo Fradkin}
\affiliation{Department of Physics, University of Illinois at
Urbana-Champaign, 1110
W.\ Green St.\ , Urbana, IL  61801-3080, USA}
\date{\today}

\newcommand{\lang} {\langle}
\newcommand{\rang} {\rangle}
\newcommand{\tphi} {{\tilde{\phi}}}
\newcommand{\ttheta} {{\tilde{\theta}}}
\newcommand{\tPi} {{\tilde{\Pi}}}
\newcommand{\tphir} {{\tilde{\phi}}_+}
\newcommand{\tphil} {{\tilde{\phi}}_-}
\newcommand{\D}{\displaystyle}
\newcommand{\sgn}{\,\mathrm{sgn}}
\newcommand{\eA}{{\EuScript{A}}}
\newcommand{\eB}{{\EuScript{B}}}
\newcommand{\eF}{{\EuScript{F}}}
\newcommand{\eH}{{\EuScript{H}}}

\begin{abstract}
A central long standing prediction of the theory of fractional quantum Hall (FQH) states that it is a topological fluid whose elementary excitations are
vortices with fractional charge and fractional statistics. Yet, the unambiguous experimental detection of this fundamental property, that the vortices have fractional statistics,
has remained an open challenge. Here we propose a  three-terminal ``T-junction'' as an experimental setup
for the direct and independent measurement of the fractional charge and statistics of fractional
quantum Hall quasiparticles via cross current noise measurements.
We present a non-equilibrium calculation of the quantum noise in the
T-junction setup for FQH Jain states. We show that the cross current
correlation (noise) can be written in a simple form, a sum of two terms, which reflects the braiding properties of the quasiparticles: the
statistics dependence captured in a factor of $\cos\theta$ in one of
two contributions.
Through analyzing these two contributions for different parameter
ranges that are experimentally relevant,
we demonstrate that the noise at finite temperature reveals signatures
of generalized exclusion principles, fractional exchange statistics
and fractional charge.
We also predict that the vortices of Laughlin states exhibit a
``bunching'' effect, while higher states in the Jain sequences 
exhibit an ``anti-bunching'' effect.

\end{abstract}
\maketitle

 \section{Introduction}
 
Bose-Einstein statistics of photons and Fermi-Dirac statistics of
electrons  hold keys to two major triumphs of quantum mechanics:
the explanation of the blackbody radiation (that started quantum mechanics) and the
periodic table.
The
spin-statistics theorem states that particles with
integer(half-integer) spins are bosons (fermions) and that the
corresponding second-quantized fields obey canonical equal time
commutation (anticommutation) relations. In three spatial
dimensions, the spin can only be integer or half-integer since the
fields should transform like an irreducible representation of the
Lorentz group $SO(3,1)$ (relativistic) or $SO(3)$
(non-relativistic). Consequently particles in three dimensional
space(3D) have either bosonic or fermionic statstics. In one
spatial dimension (1D) on the other hand, neither fermions nor
hard core bosons can experience their statistics since they cannot
go past each other. As a result, statistics is essentially
arbitrary in one spatial dimension (which in some sense can even be regarded as a matter of definition.) In particular, the excitations of (integrable) one-dimensional systems are topological solitons which have a  two-body S-matrix which acquires a phase factor upon the exchange of the positions of the solitons. In this sense, one can assign an intermediate statistics to the solitons.
However in two spatial dimensions the situation is quite different. It has long been known\cite{leinaas77,wilczek82} that in two dimensions an intermediate form of statistics, {\em fractional} statistics is possible. 
A specific quantum
mechanical construction of a particle with fractional statistics, proposed and dubbed
an {\em anyon} by \textcite{wilczek82}, consists of a particle
of charge $q$ bound to a solenoid with flux $\phi$, where $q\phi$
bears a non-integer ratio to the fundamental flux quantum
$\phi_0$. In this paper, we study the 2D setting of the quantum
Hall system as an arena for displaying such fractional statistics,
and propose a concrete experiment for measuring its effects. 

Not long after the first observation of the Fractional Quantum
Hall (FQH) effect~\cite{tsui82}, Laughlin 
proposed 
that quasiparticle(qp)s/ quasihole(qh)s of these incompressible fluids carry a {\em fractional charge}
$e^*=\pm\nu e$  determined by the precisely quantized
Hall conductance, 
$\nu=1/(2n+1)$  for an integer $n$~\cite{laughlin83} (for Laughlin states). Soon after,
it was shown that these qp's, would have fractional {\it
braiding statistics}~\cite{arovas84,halperin84}, {\it i.e.\/} the two qp
joint wave function would gain a complex valued phase factor
interpolating between $1$ (boson) and $-1$ (fermion) upon
exchange:
 \begin{equation}
\Psi({\bf r}_1, {\bf r}_2)=e^{i\theta}\Psi({\bf r}_2, {\bf r}_1).
\label{eq:exchange}
\end{equation}
with  $0<\theta<\pi$. In the anyonic picture, the statistical
angle corresponds to $\theta=\pi q\phi/\phi_0$. Laughlin qp's are a specific example with $q=\nu$. In order to compute
the statistical angle $\theta$, \textcite{arovas84} first
considered the the adiabatic process of one qp encircling another. By calculating the Berry phase
associated with this process they showed that the two qp wave
function $\Psi({\bf r}_1, {\bf r}_2)$ gains an extra ``statistical
phase''  of $e^{i2\nu\pi}$  in addition to the magnetic flux
induced Aharonov-Bohm phase upon encircling. An adiabatic exchange
of two qp's  can be achieved by moving one qp only half its way
around the other and shifting both of them rigidly to end up in
interchanged initial positions. They thus argued that the phase
factor picked up by a two qp joint wave function upon this
exchange process should be precisely half of what it is for a
round trip and hence $\theta=\nu\pi$.

 \begin{figure}
 \psfrag{r1}{${\bf r}_1$}\psfrag{r2}{${\bf r}_2$}\psfrag{t}{$t$}
 \includegraphics[width=.25\textwidth]{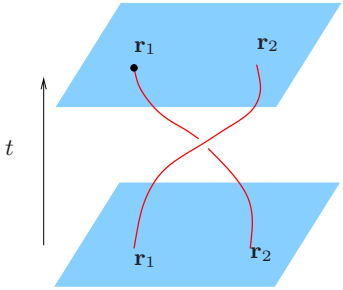}
 \caption{An adiabatic exchange between two qp's braids worlds lines representing histories of each qp. }
 \label{fig:braiding}
 \end{figure}

From a  more general standpoint, anyons are  excitations which carry a representation of the {\em braid group} (see below). This is possible both for non-relativistic particles in high magnetic fields (as we will be interested in here) as well as some field-theoretic relativistic models. As in the FQH example, the quantum mechanical amplitudes for processes involving two or more of these particles (regarded as low-energy excitations from some more complicated system) are represented by world lines which never cross (as if they had a hard-core repulsion). In $2+1$ dimensions different histories of these particles can be classified according to the topological invariants of the knots formed by their world lines (see Fig.~\ref{fig:braiding}). These topological invariants are representations of a group, the {\em braid group}. The statistical angle is one such label and it corresponds to a an Abelian (one-dimensional) representation of the braid group. Most quantum Hall states are Abelian (in this sense). However a number of non-Abelian states have been constructed ({\em e.g.\/} $\nu=5/2,12/5$ and a few others). 

Fractional (or braid) statistics is a generalization of the concept that fermions have wave functions which are odd upon exchange ({\em i.e.\/} thus they obey the Pauli Exclusion Principle)  while bosons have wave functions which are odd upon exchange. The notion of statistics is (as its name indicates) also related to the problem of counting states.
Some years ago, Haldane\cite{haldane91} introduced the concept of {\it exclusion
statistics}, which  is a
generalization of the   Pauli exclusion principle for fermions.
Exclusion statistics in finite systems is determined by how much
an addition of a particle diminishes the number of available
states for yet another addition. For fermions, the number of
states will diminish by one if  a particle is added to the system
while for bosons, the number of states would stay the same;
Haldane's idea was to consider more general possibilities
interpolating between fermions and bosons. Although the Hilbert
space counting definition of statistics and the braiding
definition of statistics coincide with one another for Laughlin
states, the two definitions are not equivalent in general. It was argued by  Haldane\cite{haldane91}, and shown explicitly by Van Elburg and Schoutens\cite{vanelburg98}, that the qp's of the FQH states also obey a form of exclusion statistics.

Fractional charge and statistics 
are fundamental properties of qp's emerging from the strongly
interacting FQH liquid. Each of the filling factors $\nu$
displaying precise quantization of fractional Hall conductance
$\sigma_{xy}=\nu e^2/h$  
represent a distinct phase
characterized by nontrivial internal order- {\it topological
order},- which is robust against arbitrary
perturbations\cite{wen-niu90,wen95}. The measurement of fractional
statistics will prove the existence of such topological order.
While there is substantial experimental evidence for the
fractional charge of these qp's and qh's~\cite{goldman95,
picciotto97,saminadayar97,reznikov99},  in particular from
two-terminal  noise experiments, similar evidence for statistical
properties or the connection between two distinct ways of
determining statistics is still lacking. The main challenge in
detection of anyons is to manipulate these ``emergent particles''
which cannot be taken outside the 2D system and to measure the
effect of statistical phase. In this paper, we show that the noise
in current fluctuations in a three-terminal geometry (a
``T-junction'') in FQH Jain states\cite{jain89} behaves as in a
Hanbury-Brown and Twiss (HBT) interferometer\cite{hbt} with clear
and independent signatures of fractional charge, fractional
statistics and exclusion statistics.

Our interest in Jain states is two fold. First, the most prominent
FQH effects which are observed lie in the Jain sequence with
$\nu=\frac{p}{2np+1}$ where $n$ and $p$ are integers. Further, as
shown in the table~\ref{table:qn}, Jain states include
``non-Laughlin states'' with fractional charge and statistical
angle that are different from the filling factor, {\it i.e.}
$e^*/e\!\neq\!\theta/\pi\!\neq\!\nu$. Therefore one can pursue
signatures of these fractional quantum numbers  independently and
predict  how each fractional quantum numbers will manifest itself
in different aspects of experimental data. Particularly, by
comparing two FQH states whose qps have same fractional charge but
different statistics one can focus on the effects of fractional
statistics.
\begin{table}[h]
\newcolumntype{Y}{>{\centering\arraybackslash$}m{1.2cm}<{$}}
\newcolumntype{C}{>{\centering\arraybackslash$}m{2.3cm}<{$}}
\renewcommand{\arraystretch}{2}
\begin{tabular}{|Y||Y|Y|C|Y|}
\hline
    & \text{boson} & p\!=\!1  & p\!>\!1& \text{fermion}\\
\hline
 e^*/e& 0,2,\cdots & \nu & \displaystyle{\frac{\nu}{p}} & 1 \\
\hline
  \theta/\pi& 0 & \nu<\displaystyle\frac{1}{2} & \left(1\!-\!\displaystyle{\frac{2n}{p}}\nu\right)>\displaystyle\frac{1}{2} & 1\\
\hline
\end{tabular}
\label{table:qn} \caption{Fractional charge $e^*$ and statistical
angle $\theta$ for qp's of FQH states at filling factor $\nu =
p/(2np\!+\!1)$ ($n$, $p$ are integers) in comparison with bosons
and fermions following Ref.[\onlinecite{lopez99}]. Alternative
descriptions predict somewhat different spectrum of qp's for
non-Laughlin states at $p\neq1$ (see Ref.[\onlinecite{wen95}]) and implications of
such differences will be discussed in the later part of this
paper.}
\end{table}
 
In this paper we present a theory of noise cross correlations of the currents in a three-terminal junction (or T-junction) which we introduced in Ref.[\onlinecite{kim05}], where we presented a summary of results of the theory. Here, we describe the theory and its conceptual and technical underpinnings in a more detailed and self-contained fashion.
The bulk of the paper consists of the description
of the cross current correlation between tunneling currents
partitioned from one edge of the proposed T-junction setup into
two others. The lowest order non-vanishing contribution to the
correlation is fourth order in tunneling and it is the quantity
which contains signatures of both fractional charge and fractional
statistics. 
In particular, by analyzing the components of the correlation in
depth, we pinpoint the role of statistics in various tunneling
processes and discuss implications of the connection between braiding statistics and exclusion statstics.

The paper is organized as follows. In Section~\ref{sec:edge} we
briefly review the effective theory for edge states developed in Ref.
[\onlinecite{lopez99}] which we will use throughout.
In Section~\ref{sec:setup}, we introduce the T-junction setup we
are proposing and we define the normalized cross current noise to
be calculated purturbatively in Section~\ref{sec:perturb}. In
Sections~\ref{sec:St} and \ref{sec:Sw}, we present the time
dependent noise and its frequency spectrum respectively. We
summarized our results and discuss its implications for experiments in
Section~\ref{sec:conclusion}. In four appendices we present a summary of the theory of the edge states that we use here (Appendix \ref{ap:boson}), of the Schwinger-Keldysh technique and conventions (Appendix \ref{ap:keldysh}), some useful identities for vertex operators (Appendix \ref{ap:vertex}), and the unitary Klein factors (Appendix \ref{ap:klein}).

\section{An effective theory for edge states}
\label{sec:edge}
In this section, we summarize the properties of edge states for
the Jain sequence and their associated quasiparticles.
 While the creation of qp excitations in the 2D
bulk of FQH systems has an associated energy gap, ({\it i.e.} FQH
liquid is incompressible), the one dimensional (1D) boundary
defined by a confining potential can support gapless
excitations\cite{wen90,wen95}. Further there is a striking
one-to-one correspondence between qp states in the bulk and at the
edge making the FQH liquid holographic\cite{wen90}.
Given that edge states comprise an effective 1D system propagating
only along the direction dictated by the magnetic field (a chiral
Luttinger liquid\cite{wen95,wen90}), they can be described in
terms of chiral bosons using the standard bosonization approach.
The edge effective Lagrangian density for Jain states, derived
from the boundary term of the fermionic Chern-Simons
theory~\cite{lopez91} by \textcite{lopez99} is given by
\begin{equation}
 \begin{split}
        {\mathcal  L}_0&=\frac{1}{4\pi\nu}\partial_x\phi_c(-\partial_t\phi_c-\partial_x\phi_c)\\
        &\qquad\quad+\lim_{v_N\rightarrow 0^+ }\frac{1}{4\pi}\partial_x\phi_N(\partial_t\phi_N+v_N\partial_x\phi_N)
\end{split}
\label{eq:L}
\end{equation}
where $\phi_c$ is a right moving charge mode whose speed is set to
be $v_c=1$ ($\phi_-$ with $g=1/\nu$ and $v=1$ case of
Appendix~\ref{ap:boson})
 and $\phi_N$ is the non-propagating charge neutral topological mode obtained as a $v_N\rightarrow 0^+$ limit of a rightmover  with $g_N=-1$.  Note that
for the purpose of regularization and to carefully keep track of
short time behavior which is crucial for ensuring the correct
statistics, we shall keep a neutral mode speed $v_N$ and 
take the limit $v_N\rightarrow0$ only at the very end.

The normal ordered vertex operator creating a quasiparticle
excitation on this edge, with fractional quantum numbers tabulated
in the Table~\ref{table:qn}, is given by~\cite{lopez99}
\begin{equation}
\psi^\dagger\propto\ \
:\!e^{-i(\frac{1}{p}\phi_c+\sqrt{1+\frac{1}{p}}\phi_N)}\!: \ \
\equiv\ \ :\!e^{-i\varphi}\!: \label{eq:qp}
\end{equation}
where we define a short hand notation $\varphi$ to represent the
appropriate linear combination of the charge mode and the
topological mode
\begin{equation}
\varphi\equiv(\frac{1}{p}\phi_c+\sqrt{1+\frac{1}{p}}\phi_N) .
\label{eq:short}
\end{equation}
By noting the equal time commutator for the new field $\varphi$ being
    \begin{equation}
    \begin{split}
    [\varphi(x,t),\varphi(x',t)]
    &=i\pi\left(\frac{\nu}{p^2}-\frac{1}{p}-1\right)\sgn(x-x')\\
    & = -i\theta\sgn(x-x')
    \end{split}
    \label{eq:varphi-comm}
    \end{equation}
the quantum numbers of $\psi^\dagger$ can be verified as follows.
The charge density operator
$j_0\equiv\frac{1}{2\pi}\partial_x\phi_c$ measures the charge of
the qp operator since $[j_0(x),\psi^\dagger(x')]\equiv
\frac{e^*}{e}\delta(x-x')\psi^\dagger(x')$. By expanding the
vertex operator Eq.~\eqref{eq:qp}, this commutator can be
calculated order by order to give
\begin{equation}
\begin{split}
[j_0(x),\psi^\dagger(x')]&=\frac{1}{2\pi}[\partial_x\phi_c(x),e^{-i\frac{1}{p}\phi_c(x')}]\\
&=-\frac{1}{2\pi}\frac{i}{p}\partial_x[\phi_c(x),\phi_c(x')]+\cdots\\
&=\frac{1}{2\pi}\frac{\nu\pi}{p}\partial_xsgn(x-x')+\cdots\\
&=\frac{\nu}{p}\delta(x-x')\psi^\dagger(x'),
\end{split}
\label{eq:charge}
\end{equation}
hence $e^*/e=\frac{\nu}{p}$. As for statistics, using the
Campbell-Baker-Hausdorff (CBH) formula for exponential of
operators $\hat{A}$, $e^{\hat{A}}e^{\hat{B}} =
e^{\hat{B}}e^{\hat{A}} e^{[\hat{A},\hat{B}]}$,
 the statistical angle of the qp operator  can be calculated via exchange of qp positions at equal time as the following
\begin{equation}
\begin{split}
\psi^\dagger(x)\psi^\dagger(x')&=\psi^\dagger(x')\psi^\dagger(x)e^{-[\varphi(x),\varphi(x')]}\\
&= \psi^\dagger(x')\psi^\dagger(x)e^{i\theta  sgn(x-x')},
\end{split}
\label{eq:angle}
\end{equation}
which confirms $\theta=(1-\frac{2n}{p}\nu)$ modulo $2\pi$.

Combining charge mode and neutral mode propagator, one can obtain the propagator for the field $\varphi$:
\begin{equation}
\begin{split}
&\langle  \varphi(x,t)\varphi(0,0)\rangle\\
&=\frac{1}{p^2}\langle 
\phi_c(x,t)\phi_c(0,0)\rangle+\left(1+\frac{1}{p}\right)\langle 
\phi_N(x,t)\phi_N(0,0)\rangle\\
&=-\frac{\nu}{p^2}\ln\left[\frac{\tau_0+i(t-x)}{\tau_0}\right]\\ &\qquad\qquad+i\lim_{v_N\rightarrow
0^+}\frac{\pi}{2}\left(1+\frac{1}{p}\right)\sgn(v_Nt-x).
\end{split}
\label{eq:varphi-prop}
\end{equation}
From Eq.~\eqref{eq:varphi-prop}, one arrives
at the following form for the  qp propagator at
zero temperature (see  Appendix~\ref{ap:vertex}):
\begin{widetext}
\begin{equation}
\begin{split}
&\langle \psi(x,t)\psi^\dagger(0,0)\rangle\\
&=e^{\langle\varphi(x,t)\varphi(0,0)\rangle}\\
&=\lim_{v_N\rightarrow 0^+}\left[\frac{\tau_0}{\tau_0+i(t-x)}\right]^{\frac{\nu}{p^2}}\
e^{-i\frac{\pi}{2}(1+\frac{1}{p})\ sgn(v_Nt-x)}.
\end{split}
\label{eq:qp-prop}
\end{equation}
\end{widetext}
Eq.~\eqref{eq:qp-prop} shows that the qp operators have the
scaling dimension $
\frac{K}{2}=\frac{\nu}{2p^2}=\frac{1}{2p(2np+1)}$.
Further, as $\tau\rightarrow 0^+$, the limit $x\rightarrow 0^-$ of
Eq.~\eqref{eq:qp-prop} becomes
\begin{equation}
\langle  \psi(0^-,t)\psi^\dagger(0,0)\rangle
 = \left|\frac{t}{\tau_0}\right|^{-K} e^{i\theta\ sgn(t)}
 \label{eq:Tqp-prop-zero}
\end{equation}
with the explicit dependence on the statistical angle emerging in
the long wavelength limit.
 This {\em equal position} propagator, which explicitly encodes information on
 statistics, plays a prominent role in subsequent point-contact tunneling calculations.
  Most importantly, the amplitude of the propagator Eq.~\eqref{eq:Tqp-prop-zero}
  is governed by the scaling dimension but the phase is solely determined by
  the {\em statistical angle}.

Since edge states are gapless excitations, they act as windows for
experimental probes to observe features of the FQH droplet. In
particular, tunneling between edges allows a viable access to
single qp properties, provided the tunneling path lie inside the
FQH liquid (qp's only exist within the liquid)\cite{chang-rmp}.In
fact, such tunneling has been successfully used in a geometry
hosting a single point contact and two edges
 for the detection of fractional charge through shot
noise measurements~\cite{picciotto97,saminadayar97,reznikov99}. In
principle, two edges would suffice for anyonic exchange as well
with qp's from each edge exchanging positions via tunneling
through the 2D FQH bulk. However, in order to detect the phase
information, the path of exchange needs to be intercepted, for
instance, by inserting a third edge and allowing qp exchange
between first two edges to go through the third edge. This third
edge thus acts at once as a stage and an ``observation deck''.
Further, if the first two edges are each adding/removing qp's to
the third edge,
 one should be able to see the generalized exclusion principle in action by probing
  the third edge.
In the next section, we propose a ``T-junction'' interferometer as
a minimal realization of such a situation.

\section{The T-junction setup}
 \label{sec:setup}
\begin{figure}[b]
\psfrag{0}{\small$0$} \psfrag{1}{\small$1$} \psfrag{2}{\small$2$}
\psfrag{V1}{\small$V_1$} \psfrag{V2}{\small$V_2$}
\psfrag{V0}{\small$V_0$}
\includegraphics[width=0.45\textwidth]{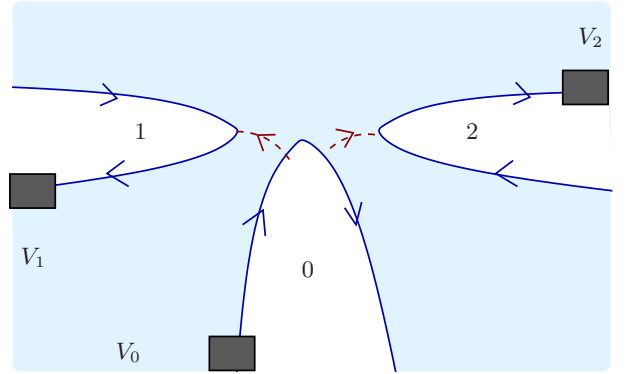}
\caption{The proposed T-junction set up.
Solid (blue) lines represent edge states where the edge state $0$
is held at potential $V$ relative to edges $1$ and $2$, {\it
i.e.\/}
 $V_0-V_1=V_0-V_2=V$. Dashed (red) lines show paths of two quasiparticle tunneling through a
 FQH liquid from  edge state $0$ to edge $1$ at $x_0\!=\!-a/2\!\equiv\!X_1$ , $x_1=0$; from edge $0$ to edge $2$ at $x_0\!=\!a/2\!\equiv\!X_2$, $x_2=0$. The direct tunneling  between edge $1$ and $2$ is turned off by setting the two edges at equal potential.
 }
 \label{fig:hbt}
\end{figure}

In our proposed T-junction interferometer of the type shown in
Fig.\ref{fig:hbt}, top gates can be used to define and bring
together three edge states $l=0,1,2$ which are separated from each
other by ohmic contacts. Two edges $l=1$ and $l=2$ each separately
form tunnel junctions with edge $0$
 and upon setting the edge $0$ at relative voltage
$V$ to the two others, qp's are driven to tunnel between edge $0$
and edges $1,2$.

Denoting charge and neutral modes for edge $l=0,1,2$ by
$\phi_c^l(x_l,t)$ and $\phi_N^l(x_l,t)$ respectively,
 each edge state $l$ in the absence of tunneling (hence the superscript $0$) can be described by
  Lagrangian densities of the form in Eq.~\eqref{eq:L}:
    \begin{equation}
        {\mathcal  L}^{(0)}_l=\frac{1}{4\pi\nu}\partial_{x_l}\phi_c^l(-\partial_t\phi_c^l-\partial_{x_l}\phi_c^l)
        +\frac{1}{4\pi}(\partial_{x_l}\phi_N^l\partial_t\phi_N^l)
    \end{equation}
where $x_l$ now represents the curvilinear abscissa along each
edge, and the total Lagrangian density becomes
    \begin{equation}
    \mathcal{L}^{(0)}=\sum_{l=0,1,2}\mathcal{L}^{(0}_l
    \label{eq:total-free-L}
    \end{equation}
Boson fields $\phi_c^l$ and $\phi_N^l$ each have the same
properties (such as commutation relations
 and propegators) as $\phi_c$ and $\phi_N$ of a single edge described in the
 Sec.~\ref{sec:edge}.
  Fields of different edges commute with each other, i.e.
    \begin{equation}
    [\phi_{c/N}^l,\phi_{c/N}^{m}]=0,\quad{\mathrm for}\;  l\neq m
    \end{equation}
for $l,m=0,1,2$.
Within each edge $l$, a vertex operator $e^{i\varphi_l}$, where
$\varphi_l$ again
 is a short hand notation  $\varphi$ defined in Eq.~\eqref{eq:short} for each edge $l$,
 describes a qp excitation $\psi_l^\dagger$ in that edge.
   However, since $\varphi_l$'s are constructed to commute between different edges,
   we need to introduce {\em unitary Klein factors} $F_l$  with the appropriate algebra, as
   shown in the Appendix~\ref{ap:klein}, to give the correct qp statistics upon exchange
    between different edges. Hence, we have
    \begin{equation}
    \psi_l^\dagger=\frac{1}{\sqrt{2\pi a_0}} F_l e^{i\left(\frac{1}{p}\phi_c^l+\sqrt{1+\frac{1}{p}}\phi_N^l\right)}\equiv \frac{1}{\sqrt{2\pi
    a_0}}F_le^{i\varphi_l},
    \label{eq:l-vertex}
    \end{equation}
where $a_0$ is the short distance cutoff and the Klein factors
obey the statistical rules
    \begin{equation}
    F_lF_m=e^{-i\alpha_{lm}}F_mF_l
    \label{eq:klein-exchange}
    \end{equation}
where $\alpha_{lm}=-\alpha_{ml}$ and
$\alpha_{02}=\alpha_{21}=-\alpha_{10}=\theta$. Notice the Klein
factors do not affect the statistics between operators on the same
edge
 since $\alpha_{ll}=0$ from the requirement $\alpha_{lm}=-\alpha_{ml}$.

Now, if we set the the distance between two tunneling points  on
edge $0$ to be a small positive real number $a$ (which we later
send to zero) and the tunneling point on edge $1, 2$ to be the
origin for the abscissa coordinate of edge $1, 2$, then qp's
tunnel at two locations , $x_0\!=\!-a/2\!\equiv\!X_1$ and
 $x_0\!=\!a/2\!\equiv\!X_2$ on edge $0$ and at $x_1\!=\!0$, $x_2\!=\!0$ on edges $1$, $2$.
The operator $\hat{V}_j(t)$ which tunnels one qp from the edge $0$
to edges $j=1,2$  at time $t$
 can then be written in the following form:
\begin{equation}
\begin{split}
\hat{V}_j^\dagger(t)&\equiv \psi_0(X_j,t)\psi_j^\dagger(0,t)\\
&=F_0F_j^{-1} e^{i\varphi_0(X_j,t)} e^{-i\varphi_j(0,t)}.
\end{split}
\label{eq:Vj}
\end{equation}
The additional term in the Lagrangian due to the presence of the
tunneling can be written as $L_{int}=L_{int,1}+L_{int,
2}$~\cite{safi01,guyon02} with tunneling term involving edges
$j=1,2$ given by
\begin{equation}
L_{int, j}(t)=-\Gamma_j \hat{V}_j^\dagger(t)+ h.c.\ ,
\label{eq:tunnel-L}
\end{equation}
where $\Gamma_j$ is the tunneling amplitude between edges $0$ and
$j\!=\!1,2$. The non-equilibrium situation of setting a constant
DC voltage bias $V$ between edges $0$ and $j$ corresponds to
substituting $\Gamma_j$ by $\Gamma_j e^{-i\omega_0 t}$ in
Eq.~\eqref{eq:tunnel-L},
  where the Josephson frequency is explicitly determined by the {\em fractional charge} to be $\omega_0\equiv e^* V/\hslash$ (the Peierls substitution~\cite{chamon95,chamon96, safi01}).
Using the Heisenberg equations of motion for the total charge
$Q_l=\int d x_l j_{0l}(x_l)$ of edge $l$  where
$j_{0l}(x_l)=\frac{1}{2\pi}\partial_x\phi_c^l$ is the charge
density operator for edge $l$, as was done by
\textcite{chamon95, chamon96} for Laughlin states, one can show that  the
tunneling current operator, $\hat{I}_j(t)$, from edge $0$ to edge
$j$ is given by
\begin{equation}
\begin{split}
\hat{I}_j(t)&=i\frac{e^*}{\hbar}\Gamma_j
(e^{i\omega_0t}\hat{V}_j^\dagger -e^{-i\omega_0t}\hat{V}_j) , \\
&\equiv i e^*\sum_{\epsilon=\pm }\epsilon\ \Gamma_j\
e^{i\epsilon\omega_0t}\ \hat{V}_j^{(\epsilon)},
\end{split}
\label{eq:current}
\end{equation}
where we assumed $\Gamma_j$ to be real and introduced the notation
$\hat{V}_j^\dagger\equiv \hat{V}_j^{(+)}$ and $\hat{V}_j \equiv
\hat{V}_j^{(-)}$, i.e.,
\begin{equation}
\hat{V}_j^{(\epsilon)}(t)=(F_0F_j^{-1})^\epsilon\
e^{i\epsilon\varphi_0(X_j, t)}\ e^{-i\epsilon\varphi_j(0,t)},
\label{eq:V-pm}
\end{equation}
 to represent the summation over hermitian conjugates in a compact form.
Notice that $\epsilon=+$ corresponds to a qp tunneling from the
edge $j$ to $0$ and $\epsilon=-$ corresponds to a reverse
direction tunneling.

We treat the non-equilibrium situation within the
Schwinger-Keldysh
formalism~\cite{keldysh64,kadanoff-baym,rammer86,chamon95,chamon96,guyon02,safi01}
for computing expectation values of operators. As a simple
example, the expectation value of the current on edge $j$ takes
the following form up to a normalization factor
\begin{equation}
\langle I_j\rangle=\frac{1}{2}\sum_{\eta=\pm} \langle T_K
\hat{I}_j(t^\eta) e^{i\int_K L_{int,j}(t_j)dt_j}\rangle_0
\label{eq:define-Ij}
\end{equation}
where $\eta$ labels the branch of the Keldysh contour:
$\eta\!=\!+$ for the forward branch
$(t\!=\!-\infty\rightarrow\infty)$ and $\eta\!=\!-$ for the
backward branch $(t\!=\!\infty\rightarrow-\infty)$. $T_K$
indicates that the operators inside brackets should be contour
ordered before taking the expectation value, and
$\langle\rangle_0$ indicates the expectation value with respect to
the free action $S_0=\int_K\!dt\int d^3 x \mathcal{L}_0$ where the
free Lagrangian density $\mathcal{L}_0$ was given in
Eq.~\eqref{eq:total-free-L}.

Our goal is to identify a measurable quantity that distinctly
captures the role of the statistical angle $\theta$ which features
in the single qp operator shown in Eq.~\eqref{eq:angle}. The
normalized cross correlation (noise) $S(t)$ between tunneling
current fluctuations
 ($\Delta I_l=I_l-\langle I_l \rangle$):
    \begin{equation}
    S(t-t')
    \equiv\frac{ \langle \Delta I_1(t)\Delta I_2(t')\rangle}
    {\langle I_1\rangle\langle I_2\rangle}
        \label{eq:define-St}
    \end{equation}
turns out to be the simplest quantity that can exhibit the subtle
signatures of statistics and distinguish them from those of
fractional charge. In fact, in contrast to shot noise, this cross
current correlation $S(t-t')$ carries statistical information even
at the lowest non-trivial order, which we calculate perturbatively
in the next section.

\section{Perturbative calculation}
 \label{sec:perturb}
 In this section, we perform a detailed analysis of the normalized
 cross correlation at fourth (lowest non-vanishing) order in
 tunneling. We calculate the two point correlation and four point correlation
 functions required to calculate the cross correlation in terms of
 edge state properties. We analyze the detailed behavior of the
 cross correlation over different Keldysh time domains and
 pinpoint the effects of statistics, specifically drawing the connection between braiding statistics
 and exclusion statistics. We show that cross correlation can be expressed in terms
  of two universal scaling functions and that the
 statistical dependence ultimately takes on a remarkably simple form.

It must be remarked that since the perturbation
introduced in Eq.~\eqref{eq:tunnel-L} is relevant in the
renormalization group sense, perturbation theory is valid only
under certain conditions. Specifically, perturbation holds as long
as energies involved, such as voltage and temperature, are higher
than a cross-over energy scale proportional to  $\Gamma^{\frac{1}{1-K}}$.
For energies much lower than this cross-over scale, qp tunneling
between edges becomes large enough to break the Hall droplet into
smaller disconnected regions and one needs an appropriate dual picture to properly address the strong coupling fixed point.
For observing the effects
of statistics as prescribed by the correlation calculated here, it
is imperative that the system remains in the perturbative regime.

 The un-normalized cross correlation $\langle I_1\rangle\langle I_2\rangle S(t-t')=\langle \Delta I_1(t)\Delta I_2(t')\rangle$, where $\langle\rangle$ represents expectation value with respect to the full Lagrangian in the presence of tunneling events, can be written in terms of expectation values with respect to the free theory  $\langle\rangle_0$ as
     \begin{widetext}
    \begin{equation}
    \begin{split}
    \langle I_1\rangle\langle I_2\rangle S(t-t')&=\frac{1}{4}\sum_{\eta,\eta'}\langle T_K\hat{I}_1(t^\eta)\hat{I}_2(t'^{\eta'})\ e^{i\int_K L_{int, 1}(t_1)dt_1+i\int_K L_{int, 2}(t_2)dt_2}\rangle_0\\
    &\qquad\qquad - \frac{1}{4}\sum_{\eta,\eta'}\langle T_K \hat{I}_1(t^\eta) e^{i\int_K L_{int, 1}(t_1)dt_1}\rangle_0\langle T_K \hat{I}_2(t'^{\eta'}) e^{i\int_K L_{int, 2}(t_2)dt_2}\rangle_0.
    \end{split}
    \label{eq:St}
    \end{equation}
    \end{widetext}

By expanding the exponentials in Eq.~\eqref{eq:St}, we can
 calculate the cross correlation perturbatively in the tunneling amplitude $\Gamma_j$. Since the
  tunneling operator is relevant in the RG sense (its scaling dimension is less than $1$), the
   perturbation theory will break down in the infrared (IR) limit. However, the finite temperature
    in our case provides a natural IR cutoff making this perturvative calculation meaningful. In fact,
     we will show in the following sections that  the strongest statistical dependence is obtained at
      temperatures comparable to the bias voltage.

Clearly the lowest non-vanishing term in the perturbation
expansion is of order $(\Gamma_1\Gamma_2)^2$. Using the expression
for the tunneling current operator in Eq.~\eqref{eq:current},  the
first term of Eq.~\eqref{eq:St} to this order can be written as
    \begin{widetext}
    \begin{equation}
    \begin{split}
    \langle I_1(t)I_2(t')\rangle^{(2)}
    &\equiv (-i)^2(ie^*)^2\!\frac{1}{4}\sum_{\eta,\eta',\eta_1,\eta_2}\!\sum_{\epsilon,\epsilon'}\epsilon\epsilon'\eta_1\eta_2
    |\Gamma_1\Gamma_2|^2\int\!\! dt_1dt_2 e^{i\epsilon\omega_0(t-t_1)+i\epsilon'\omega_0(t'-t_2)}\\
    &\qquad\qquad\qquad \times\langle T_K \hat{V}_1^{(\epsilon)}(t^\eta) \hat{V}_2^{(\epsilon')}(t'^{\eta'}) \hat{V}_1^{(-\epsilon)}(t_1^{\eta_1}) \hat{V}_2^{(-\epsilon')}(t_2^{\eta_2})\rangle_0
    \end{split}
    \label{eq:I1I2-second}
    \end{equation}
    \end{widetext}
where we used the fact that the Keldysh contour time-integral in
the exponent can be represented using the contour branch index
$\eta'$
    \begin{equation}
    \int_K dt = \sum_{\eta'=\pm}\eta'\int dt
    \end{equation}
and the charge neutrality condition discussed in the
Appendix~\ref{ap:vertex} which enforces $\epsilon_1=-\epsilon$ and
$\epsilon_2=-\epsilon'$. Similarly, the expectation value of the
current for the second term of Eq.~\eqref{eq:St} becomes
    \begin{multline}
    \langle \hat{I}_1\rangle^{(2)}=
        \frac{1}{2}\sum_{\epsilon=\pm1}\sum_{\eta,\eta_1}(-i)(ie^*)\epsilon\eta_1\int dt_1\bigg[\\
        |\Gamma_1|^2 e^{i\epsilon\omega_0(t-t_1)}
        \langle T_K \hat{V}_1^{(\epsilon)}(t^\eta)\hat{V}_1^{(-\epsilon)}(t_1^{\eta_1})\rangle_0\bigg]
    \label{eq:I-second}
    \end{multline}
Putting Eq.~\eqref{eq:I1I2-second} and Eq.~\eqref{eq:I-second}
together, we arrive at the following expression for the lowest
non-vanishing term of the un-normalized cross current correlation
    \begin{widetext}
    \begin{equation}
    \begin{split}
    \langle I_1\rangle\langle I_2\rangle S^{(2)}(t-t')
    &\; =\frac{1}{4}(e^*)^2\!\sum_{\eta,\eta',\eta_1,\eta_2}\!\!\sum_{\epsilon,\epsilon'}\epsilon\epsilon'\eta_1\eta_2
    |\Gamma_1\Gamma_2|^2\int\!\! dt_1dt_2 e^{i\epsilon\omega_0(t-t_1)+i\epsilon'\omega_0(t'-t_2)}\\
    &\quad \times \left[ \langle T_K \hat{V}_1^{(\epsilon)}(t^\eta) \hat{V}_2^{(\epsilon')}(t'^{\eta'}) \hat{V}_1^{(-\epsilon)}(t_1^{\eta_1}) \hat{V}_2^{(-\epsilon')}(t_2^{\eta_2})\rangle_0\right.\\
    &\left.\qquad\qquad\qquad\qquad-\langle T_K \hat{V}_1^{(\epsilon)}(t^\eta)  \hat{V}_1^{(-\epsilon)}(t_1^{\eta_1}) \rangle_0\langle T_K\hat{V}_2^{(\epsilon')}(t'^{\eta'}) \hat{V}_2^{(-\epsilon')}(t_2^{\eta_2})\rangle_0
    \right].
    \end{split}
    \label{eq:S-second}
    \end{equation}
    \end{widetext}
We now evaluate the two point function and the four point
function, and then combine them to obtain a simplified formula for
$\langle I_1\rangle\langle I_2\rangle S^{(2)}(t-t') $ from
Eq.~\eqref{eq:S-second}.

\subsection{The two point function}
\label{subsec:twopoint}
Using Eq.~\eqref{eq:Vj} and the result of Appendix~\ref{ap:vertex}
to calculate the chiral
 boson vertex correlation function, we find
    \begin{align}
    \langle &T_K \hat{V}_1^{(\epsilon)}(t^\eta)\hat{V}_1^{(-\epsilon)}(t_1^{\eta_1})\rangle_0\nonumber\\
    &=(F_0F_1^{-1})^{\epsilon}(F_0F_1^{-1})^{-\epsilon}\langle T_K e^{i\epsilon\varphi_0(X_1, t^\eta)} e^{-i\epsilon\varphi_0(X_1, t^{\eta_1}_1)}\rangle\nonumber\\
    &\qquad \times\langle T_K e^{i\epsilon\varphi_1(0, t^\eta)}
         e^{-i\epsilon\varphi_1(0, t^{\eta_1}_1)}\rangle\\
    &=\exp[ \tilde{G}_{\eta,\eta_1}(0, t\!-\!t_1)]\exp[ \tilde{G}_{\eta,\eta_1}(0, t\!-\!t_1)]
    \label{eq:two-tunneling}
    \end{align}
where $ \tilde{G}_{\eta,\eta'}(x\!-\!x', t\!-\!t')\equiv\langle
T_K \varphi_l(x,t^\eta)\varphi_l(x',t'^{\eta'})\rangle_0$ is the
Keldysh ordered and regulated $\varphi_l$  propagator for all $l$
which will be discussed below in
 detail for both zero temperature and finite temperatures.  Here we also used the fact that $\varphi_l$'s are
  independent from each other for different $l$'s. Notice that the Klein factor contribution simply
  becomes an identity for a two point function of the tunneling operator $\hat{V}_1$. Hence, Klein factors
  are not necessary in calculations involving two point functions alone, and in fact, this is the reason that
  the lowest order contribution to shot noise does not contain statistical information.
However, as we show below, Klein factors have non-trivial
contributions in the four point function which appears in the
first term of Eq.~\eqref{eq:St}.

We now focus on the form of $G_{\eta,\eta'}(-a, t-t')$ in the
limit $a\rightarrow 0^+$.
 From the detailed analysis of Appendix~\ref{ap:keldysh}, the contour ordered propagators for
 $\phi_c$ and $\phi_N$ takes the following forms  at finite
 temperatures:
    \begin{align}
    \langle &T_K \phi_c(x, t^\eta)\phi_c(x', t'^{\eta'})\rangle \nonumber\\
    &= -\nu\ln\left[ \frac{\sin\!\frac{\pi}{\beta}\!    [\tau_0+i\chi_{\eta,\eta'}(t-t')\{(t\!-\!t')\!-\!(x-x')\}]}{\frac{\pi\tau_0}{\beta}}\right]\\
    \langle &T_K \phi_N(x, t^\eta)\phi_N(x', t'^{\eta'})\rangle\nonumber\\
     &=  \lim_{v_N\rightarrow 0^+} -i\frac{\pi}{2}\chi_{\eta,\eta'} \sgn[v_N(t-t')-(x-x')].
    \end{align}
Hence the contour ordered propagator for $\varphi$ represented by
$\tilde{G}$ in Eq.~\eqref{eq:two-tunneling} and
Eq.~\eqref{eq:four-tunneling} can be written as
\begin{widetext}
    \begin{eqnarray}
  &&  \tilde{G}_{\eta, \eta'}(-a, t-t')=
   -\frac{\nu}{p^2}\ln\left[ \frac{\sin\!\frac{\pi}{\beta}\![\tau_0+i\chi_{\eta,\eta'}(t-t')( t\!-\!t'+ a)]}{\frac{\pi\tau_0}{\beta}}\right]
     +\lim_{v_N\rightarrow 0^+}i\frac{\pi}{2}\left(\frac{1}{p}+1\right) \chi_{\eta, \eta'}(t-t') \sgn[v_N(t-t')+a] \nonumber \\
   &&
    \label{eq:keldyshG-small-a}
    \end{eqnarray}
    \end{widetext}
For the case of interest,i.e., $a\rightarrow 0^+$ in the limit
$\tau_0\rightarrow 0^+$, the above takes the following form (See
Appendix~\ref{ap:keldysh}):
    \begin{multline}
    \tilde{G}_{\eta, \eta'}(0^-, t-t') = \ln C(t-t';\, T,K)  \\+ i\frac{\theta}{2}\chi_{\eta,\eta'}(t-t')\sgn(t-t').
    \label{eq:keldyshG-zero-a}
    \end{multline}
where we have introduced a notation for the amplitude of the qp
propagator that depends on the temperature $T$ and the scaling
dimension $K=\nu/p^{2}$:
    \begin{equation}
    C(t;\,T,K)\equiv \frac{(\pi\tau_0)^{K}}{\left|\sinh \pi k_BT
    t\right|^K},
    \label{eq:def-C}
    \end{equation}
and defined
    \begin{equation}
    \chi_{\eta,\eta'}(t-t') \equiv \left(\frac{\eta+\eta'}{2}\sgn(t-t')-\frac{\eta-\eta'}{2}\right) .
    \label{eq:chi-text}
    \end{equation}

We can now calculate the expectation value of the tunneling
current $\langle I_1\rangle$ of Eq.~\eqref{eq:I-second} to order
$O(\Gamma_1^2)$ using Eq.~\eqref{eq:two-tunneling}  and
Eq.~\eqref{eq:keldyshG-small-a},
    \begin{equation}
    \langle I_1(t^\eta)\rangle^{(2)} =e^*|\Gamma_1|^2\sum_\epsilon \epsilon\int dt_1e^{i\epsilon\omega_0(t-t_1)}\Upsilon(\theta)C(t-t_1)^2
    \label{eq:I1}
    \end{equation}
with the statistical factor
    \begin{equation}
    \Upsilon(\theta) = \left(\sum_{\eta_1}\eta_1e^{i\theta\chi_{\eta,\eta_1}(t-t_1)\sgn(t-t_1)}\right)
    = 2i\sin\theta
    \end{equation}
Notice that the resulting $\langle I_1(t^\eta)\rangle$ is
independent of $\eta$ and is a constant independent of $t$, as
applicable for a steady state current.

\subsection{The four point function}
\label{subsec:fourpoint}

Following a procedure similar to the one that led up to
Eq.~\eqref{eq:two-tunneling}, we find that the four point function
becomes
    \begin{widetext}
    \begin{align}
     &\langle T_K \hat{V}_1^{(\epsilon)}(t^\eta) \hat{V}_2^{(\epsilon')}(t'^{\eta'}) \hat{V}_1^{(-\epsilon)}(t_1^{\eta_1}) \hat{V}_2^{(-\epsilon')}(t_2^{\eta_2})\rangle_0\nonumber\\
     &=\langle T_K e^{i\epsilon\varphi_0(X_1, t^\eta)} e^{i\epsilon'\varphi_0(X_2, t'^{\eta'})} e^{-i\epsilon\varphi_0(X_1, t_1^{\eta_1})} e^{-i\epsilon'\varphi_0(X_2, t_2^{\eta_2})} \rangle_0
      \langle T_K e^{i\epsilon\varphi_1(0,t^\eta)} e^{-i\epsilon\varphi_1(0,t_1^{\eta_1})} \rangle_0
     \nonumber\\ &\qquad\times
    \langle T_K e^{i\epsilon'\varphi_2(0,t'^{\eta'})} e^{-i\epsilon'\varphi_2(0,t_2^{\eta_2})}\rangle_0
    \langle T_K (F_0F_1^{-1})^\epsilon(F_0F_2^{-1})^{\epsilon'}(F_0F_1^{-1})^{-\epsilon}(F_0F_2^{-1})^{-\epsilon'}\rangle_0\\
     &= e^{[  \tilde{G}_{\eta,\eta_1}(0, t\!-\!t_1)+ \tilde{G}_{\eta',\eta_2}(0, t'\!-\!t_2)+\epsilon\epsilon'  \tilde{G}_{\eta,\eta_2}(-a, t\!-\!t_2)+\epsilon\epsilon' \tilde{G}_{\eta'\eta_1}(-a, t'\!-\!t_1)-\epsilon\epsilon' \tilde{G}_{\eta\eta'}(a, t-t')-\epsilon\epsilon'  \tilde{G}_{\eta_1\eta_2}(-a, t_1\!-\!t_2)]}\nonumber\\
     &\qquad\qquad \times e^{ \tilde{G}_{\eta\eta_1}(0, t\!-\!t_1)} e^{ \tilde{G}_{\eta'\eta_2}(0, t'\!-\!t_2) }
     \langle T_K (F_0F_1^{-1})^\epsilon(F_0F_2^{-1})^{\epsilon'}(F_0F_1^{-1})^{-\epsilon}(F_0F_2^{-1})^{-\epsilon'}\rangle_0\\
     &=e^{2[ \tilde{G}_{\eta\eta_1}(0, t\!-\!t_1) +  \tilde{G}_{\eta'\eta_2}(0, t'\!-\!t_2)]}
     \frac{ e^{\epsilon\epsilon' [ \tilde{G}_{\eta,\eta_2}(-a, t\!-\!t_2)+ \tilde{G}_{\eta'\eta_1}(a, t'\!-\!t_1)]}}
     { e^{\epsilon\epsilon'[ \tilde{G}_{\eta\eta'}(-a, t\!-\!t')+ \tilde{G}_{\eta_1\eta_2}(-a, t_1\!-\!t_2)]}}
     \langle T_K (F_0F_1^{-1})^\epsilon(F_0F_2^{-1})^{\epsilon'}(F_0F_1^{-1})^{-\epsilon}(F_0F_2^{-1})^{-\epsilon'}\rangle_0
     \label{eq:four-tunneling}
    \end{align}
    \end{widetext}
where we used Eq.~\eqref{eq:neutral-vertex-corr} for the case
$N~\!=\!~4$ to calculate the four vertex correlator for the second
equality and used the fact that $X_1-X_2= -a$.

The factor involving four sets of Klein factors  of the form
$(F_0F_j^{-1})^{\pm}$ in the Eq.~\eqref{eq:four-tunneling} has to
be appropriately rearranged through proper exchange rules,
Eqs.~\eqref{eq:Flm-exchange} and \eqref{eq:Flm-conj-exchange}, as
the tunneling operators $\hat{V}_j(t^\eta)$'s get rearranged for
Keldysh contour ordering. Since contour ordering depends on time
arguments of the tunneling operators, Klein factors effectively
become endowed with dynamics in the sense that where each of these
four times fall on the contour determines whether Klein factors
associated with tunneling at each time have to be moved or
not\cite{guyon02}. In order to rearrange four sets of combined
Klein factors of the form $(F_0F_j^{-1})^\epsilon(t^\eta)$
involved in Eq.~\eqref{eq:four-tunneling}, we have to take  six
different possible pairs and contour order within each pair. This
process can be thought of as a full ``contraction'' where pairs of
Klein factors can be treated akin to propagators.

These contour ordered  Klein factor ``propagators'' were
calculated in
 Appendix~\ref{ap:klein} to be
    \begin{align}
    \langle &T_K  (F_0F_1^{-1})^\epsilon(t^\eta)(F_0F_2^{-1})^{\epsilon'}(t'^{\eta'})\rangle_0
     \nonumber\\
    &= e^{i\epsilon\epsilon'\frac{\theta}{2}\sgn(X_1-X_2) \chi_{\eta,\eta'}(t-t')} = e^{-i\epsilon\epsilon'\frac{\theta}{2}\chi_{\eta,\eta'}(t-t')}\\
    \langle& T_K  (F_0F_2^{-1})^\epsilon(t^\eta)(F_0F_1^{-1})^{\epsilon'}(t'^{\eta'})\rangle_0
    \nonumber\\
    &= e^{i\epsilon\epsilon'\frac{\theta}{2}\sgn(X_2-X_1) \chi_{\eta,\eta'}(t-t')} = e^{-i\epsilon\epsilon'\frac{\theta}{2}\chi_{\eta',\eta}(t'-t)}\\
    \label{eq:chi1}
    \end{align}
where $\chi_{\eta,\eta'}(t)$ is defined in
Eq.~\eqref{eq:chi-text}. Further using the fact that $F_0F_j^{-1}$
commutes with itself gives
\begin{equation}
\langle T_K
(F_0F_j^{-1})^{\epsilon}(t^\eta)(F_0F_j^{-1})^{-\epsilon'}(t'^{\eta'})\rangle_0=1.
\label{eq:chi2}
\end{equation}
Expressing the Klein factor contribution in
Eq.~\eqref{eq:four-tunneling} in terms of the six possible pairs
and using their forms given by Eq.~\eqref{eq:chi1} and
Eq.~\eqref{eq:chi2} simplifies it to the form
    \begin{equation}
    \frac{e^{-i\epsilon\epsilon'\frac{\theta}{2}\chi_{\eta,\eta_2}(t-t_2)}
    e^{-i\epsilon\epsilon'\frac{\theta}{2}\chi_{\eta_1,\eta'}(t_1-t')} }
    { e^{-i\epsilon\epsilon'\frac{\theta}{2}\chi_{\eta,\eta'}(t-t')}
     e^{-i\epsilon\epsilon'\frac{\theta}{2}\chi_{\eta_1,\eta_2}(t_1-t_2)}}.
    \end{equation}
Hence, we can rewrite the four point function of
Eq.~\eqref{eq:four-tunneling} as the following:
    \begin{widetext}
    \begin{equation}
    \begin{split}
     &\langle T_K \hat{V}_1^{(\epsilon)}(t^\eta) \hat{V}_2^{(\epsilon')}(t'^{\eta'}) \hat{V}_1^{(-\epsilon)}(t_1^{\eta_1}) \hat{V}_2^{(-\epsilon')}(t_2^{\eta_2})\rangle_0\\
    &=e^{2[ \tilde{G}_{\eta\eta_1}(0,\ t-t_1) +  \tilde{G}_{\eta'\eta_2}(0,\ t'-t_2)]}
     \frac{ e^{\epsilon\epsilon' [ \tilde{G}_{\eta,\eta_2}(-a,\ t-t_2)-i\frac{\theta}{2}\chi_{\eta,\eta_2}(t-t_2)    + \tilde{G}_{\eta_1,\eta'}(-a, t_1-t')-i\frac{\theta}{2}\chi_{\eta_1,\eta'}(t_1-t')  ] }}
     { e^{\epsilon\epsilon'[ \tilde{G}_{\eta,\eta'}(-a,\ t-t')-i\frac{\theta}{2}\chi_{\eta,\eta'}(t-t')    + \tilde{G}_{\eta_1,\eta_2}(-a,\ t_1-t_2)-i\frac{\theta}{2}\chi_{\eta_1,\eta_2}(t_1-t_2)      ]}}
     \end{split}
     \label{eq:four-combined}
    \end{equation}
    \end{widetext}
where we  used the fact that $G_{\eta, \eta'} (x-x', t-t' ) =
G_{\eta', \eta}(x'-x, t'-t)$, as it is pointed out in the
Appendix~\ref{ap:keldysh}, to replace $G_{\eta',\eta_1}(a,\
t'-t_1)$ by $G_{\eta_1,\eta'}(-a,\ t_1-t')$.

 Furthermore, observing the following,
    \begin{align}
    &\tilde{G}_{\eta, \eta'}(0^-,\ t-t') -i\frac{\theta}{2}\chi_{\eta,\eta'}(t-t')\nonumber\\
    &= -K\ln\left|\frac{\sinh\left(\frac{\pi (t-t')}{\beta}\right)}{\frac{\pi\tau_0}{\beta}}\right|
    +i\frac{\theta}{2}\eta(1-\sgn(t-t'))\\
    &= \tilde{G}_{\eta, -\eta}(0^-,\ t-t'),
    \end{align}
where we have used Eq.~\eqref{eq:keldyshG-zero-a} and
Eq.~\eqref{eq:chi-text}, the four point function of
Eq.~\eqref{eq:four-combined} further simplifies to the final form
    \begin{align}
    &\langle T_K \hat{V}_1^{(\epsilon)}(t^\eta) \hat{V}_2^{(\epsilon')}(t'^{\eta'})
        \hat{V}_1^{(-\epsilon)}(t_1^{\eta_1}) \hat{V}_2^{(-\epsilon')}(t_2^{\eta_2})\rangle_0\nonumber\\
    &=e^{2[ \tilde{G}_{\eta\eta_1}(0,\ t-t_1) +  \tilde{G}_{\eta'\eta_2}(0,\ t'-t_2)]}\nonumber\\
    & \qquad\times\frac{ e^{\epsilon\epsilon' [ \tilde{G}_{\eta,-\eta}(0^-,\ t-t_2) + \tilde{G}_{\eta_1,-\eta_1}(0^-, t_1-t')] }}
     { e^{\epsilon\epsilon'[ \tilde{G}_{\eta,-\eta}(0^-,\ t-t') + \tilde{G}_{\eta_1,-\eta_1}(0^-,\ t_1-t_2)   ]}} \\
     &\equiv\frac{(\pi\tau_0)^{2K}}{ \left| \sinh\frac{\pi(t-t_1)}{\beta} \right|^{2K} }
     \frac{(\pi\tau_0)^{2K}}{ \left| \sinh\frac{\pi(t'-t_2)}{\beta} \right|^{2K} }\nonumber\\
    &\qquad\times\frac{   \left| \sinh\frac{\pi(t_1-t_2)}{\beta} \right|^{\tilde{\epsilon}K}  \left| \sinh\frac{\pi(t-t')}{\beta} \right|^{\tilde{\epsilon}K}  }{   \left| \sinh\frac{\pi(t_1-t')}{\beta} \right|^{\tilde{\epsilon}K}  \left| \sinh\frac{\pi(t-t_2)}{\beta} \right|^{\tilde{\epsilon}K} } e^{i\Phi^{\eta,\eta',\eta_1,\eta_2}_{\tilde{\epsilon}}(t,t',t_1,t_2)}
    \label{eq:four-amp-phase}
    \end{align}
    where we defined $\tilde{\epsilon}\equiv\epsilon\epsilon'$ noting that Eq.~\eqref{eq:four-combined} depends only on the product $\epsilon\epsilon'$ and collected all contributions to the phase factor by  defining the following function
    \begin{align}
    &\Phi^{\eta,\eta',\eta_1,\eta_2}_{\tilde{\epsilon}}(t,t',t_1,t_2) \nonumber\\
    &\equiv \theta\chi_{\eta,\eta_1}(t-t_1)+\theta\chi_{\eta',\eta_2}(t'-t_2)\nonumber\\
    &\qquad-\tilde{\epsilon}\frac{\theta}{2}\Big[\eta\sgn(t-t_2)-\eta\sgn(t-t')\Big] \nonumber\\
    &\qquad-\tilde{\epsilon}\frac{\theta}{2}\Big[\eta_1\sgn(t_1-t')-\eta_1(t_1-t_2)\Big]\\
    &=-\eta\frac{\theta}{2}\{\sgn(t-t_1)+\tilde{\epsilon}\sgn(t-t_2)\nonumber\\
    &\qquad-\tilde{\epsilon}\sgn(t-t')-1\}+\eta'\frac{\theta}{2}\{1-\sgn(t'-t_2)\} \nonumber\\
    &\qquad -\eta_1\frac{\theta}{2}\{-\sgn(t-t_1)+\tilde{\epsilon}\sgn(t_1-t')\nonumber\\
    &\qquad-\tilde{\epsilon}\sgn(t_1-t_2)-1\} +\eta_2\frac{\theta}{2}\{1+\sgn(t'-t_2)\}.
    \label{eq:Phi}
    \end{align}

    \begin{figure}[h]
    \psfrag{e+}{$S$}
    \psfrag{e-}{$O$}
    \psfrag{t1}{\small{$t_1$}}
    \psfrag{t2}{\small{$t_2$}}
    \subfigure[]{\includegraphics[width=0.2\textwidth]{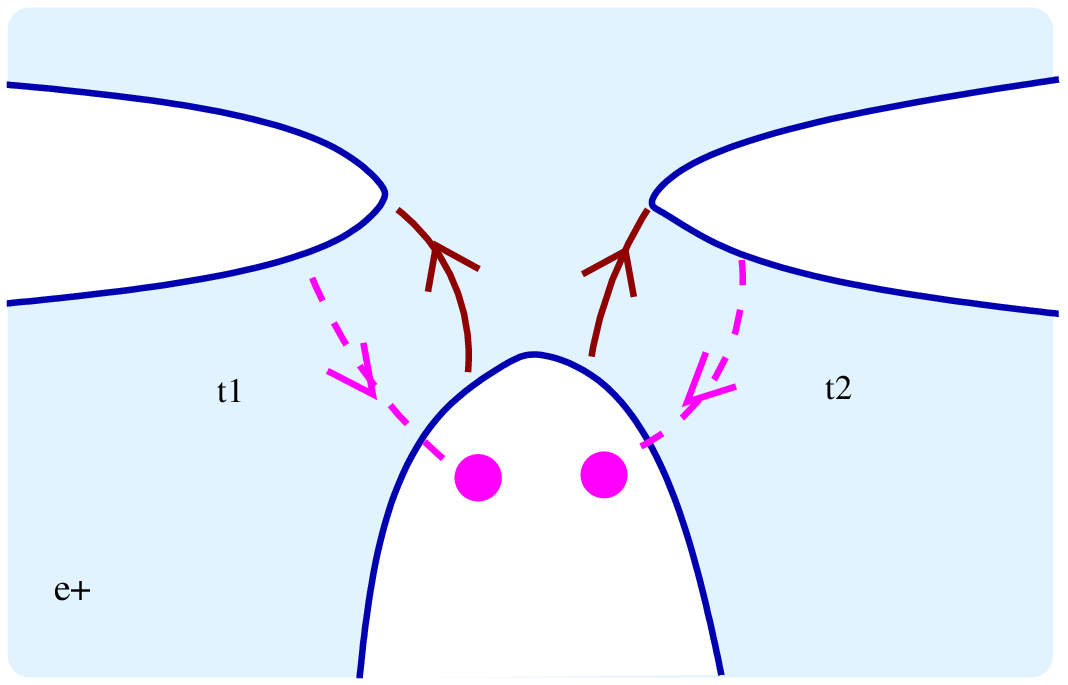}} \;
    \subfigure[]{\includegraphics[width=0.2\textwidth]{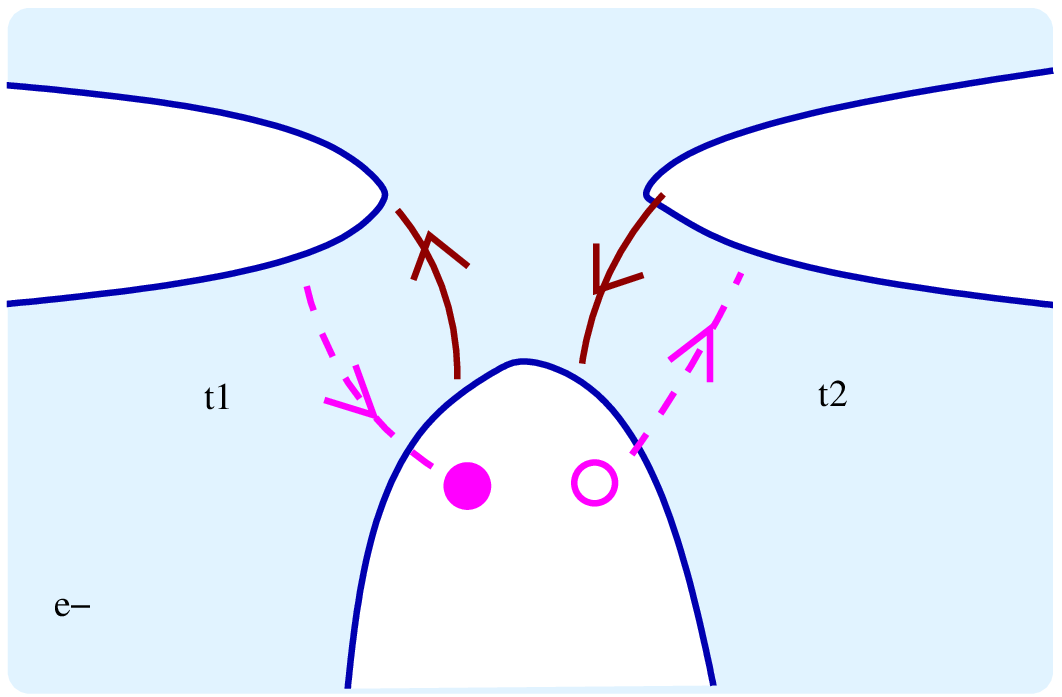}}
    \caption{Virtual tunneling events taking place at all times $t_1$ and $t_2$ allowed by causality contribute to the noise (tunneling current cross-correlation) $S(t)$ in the lowest non-vanishing order.  Solid (red) lines show paths of two quasiparticles tunneling through a FQH liquid from one edge state to two others (consitituting $I_1$ and $I_2$) and dashed (magenta) lines represent virtual tunneling events. Depending on the relative orientation of the currents (or equivalently of the virtual processes) there are two cases: (a) the case $S$ in which two currents are in the same orientation ($\tilde{\epsilon} =+1$ in the text) and (b) the case $O$ in which two currents are in the opposite direction ($\tilde{\epsilon} = -1$ in the text). }
    \label{fig:SO}
    \end{figure}

Note that  $\tilde{\epsilon} \equiv\epsilon\epsilon'$ depends only
on the relative directions of tunneling,
 that is $\tilde{\epsilon}= +1$ for correlations between contributions to $I_1$ and $I_2$ in the same
  orientation ($\epsilon=\epsilon'$) and $\tilde{\epsilon}=-1$ for correlations between
   contributions to $I_1$ and $I_2$ in the opposite orientation ($\epsilon=-\epsilon'$).
   Each case of $\tilde{\epsilon}=\pm$ is illustrated in Fig.~\ref{fig:SO} labeled $S/O$ respectively. The fact that the phase factor $\Phi$ depends on these relative orientations allows us to understand the statistical angle dependent contribution to the noise in connection with {\em exclusion statistics}~\cite{haldane91} as we will discuss in the next section.  Another important observation to be made in Eq.~\eqref{eq:Phi} is that the time dependence of the phase factor is such that it only depends on the sign of various time differences. Hence as we integrate over virtual times  $t_1$ and $t_2$, to evaluate $\langle I_1(t)I_2(0)\rangle$ according to Eq.~\eqref{eq:I1I2-second},
the domain of integration in $(t_1,t_2)$ space can be divided into
sub-domains as shown in Fig.~\ref{fig:map}  and the phase factors
for a given set of Keldysh branch indices $\eta$'s and relative
orientation of tunneling $\tilde{\epsilon}$ may be computed. Since
integration is additive, we can consider each domain's
contribution separately and later add them up to unravel the
effect of the phase factor. A simple analysis to follow shows that
large parts of the $(t_1,t_2)$ space do not contribute to the
integral due to vanishing summation over $\eta$'s. The remaining
domains can then be assigned different values of overall constant
from the summation over the phase factor.

\subsection{The normalized noise}
 \label{subsec:noise}

Putting Eq.~\eqref{eq:four-amp-phase} back into
Eq.~\eqref{eq:I1I2-second} for the $O(\Gamma_1^2\Gamma_2^2)$ term
of the current-current correlation and setting $t'=0$  (the
expression has time translational invariance) and taking $t>0$, we
obtain for $\langle I_1(t)I_2(0)\rangle^{(2)}$
    \begin{widetext}
    \begin{equation}
    \begin{split}
    &\langle I_1(t)I_2(0)\rangle^{(2)}\\
    &= (-i)^2(ie^*)^2\!\! \sum_{\eta_1,\eta_2}\sum_{\epsilon,\epsilon'}\epsilon\epsilon'\eta_1\eta_2
    |\Gamma_1\Gamma_2|^2\int\!\! dt_1dt_2\;e^{i\epsilon\omega_0(t-t_1)-i\epsilon'\omega_0t_2} \;e^{i\Phi^{\eta,\eta',\eta_1,\eta_2}_{\tilde{\epsilon}}(t,0,t_1,t_2)}\\
    &\qquad \qquad\times
    \frac{(\pi\tau_0)^{2K}}{ \left| \sinh\frac{\pi(t-t_1)}{\beta} \right|^{2K} }
     \frac{(\pi\tau_0)^{2K}}{ \left| \sinh\frac{\pi t_2}{\beta} \right|^{2K} }
     \left( \frac{   \left| \sinh\frac{\pi(t_1-t_2)}{\beta} \right|^{K}  \left| \sinh\frac{\pi t}{\beta} \right|^{K}  }{   \left| \sinh\frac{\pi t_1}{\beta} \right|^{K}  \left| \sinh\frac{\pi(t-t_2)}{\beta} \right|^{K} } \right)^{\epsilon\epsilon'}\\
        &=2{e^*}^2|\Gamma_1\Gamma_2|^2\sum_{\tilde{\epsilon}=\pm}\int\!\! dt_1dt_2\;\tilde{\epsilon}\;\cos[\omega_0(t-t_1-\tilde{\epsilon}t_2)]\\
    &\quad\times \left(\sum_{\eta_1,\eta_2=\pm} \eta_1\eta_2 e^{i\Phi^{\eta,\eta_1,\eta_2}_{\tilde{\epsilon}}(t,0,t_1,t_2)}\right)
    \frac{(\pi\tau_0)^{2K}}{ \left| \sinh\frac{\pi(t-t_1)}{\beta} \right|^{2K} }
     \frac{(\pi\tau_0)^{2K}}{ \left| \sinh\frac{\pi t_2}{\beta} \right|^{2K} }
    \left( \frac{   \left| \sinh\frac{\pi(t_1-t_2)}{\beta} \right|^{K}  \left| \sinh\frac{\pi t}{\beta} \right|^{K}  }{   \left| \sinh\frac{\pi t_1}{\beta} \right|^{K}  \left| \sinh\frac{\pi(t-t_2)}{\beta} \right|^{K} } \right)^{\tilde{\epsilon}}
    \end{split}
    \label{eq:I1I2-Phi}
    \end{equation}
    \end{widetext}
where we separated the summation over Keldysh branch indices
$\sum_{\eta,\eta',\eta_1,\eta_2}$ from the rest of the integrand using
the fact that Keldysh branch index dependence enters the
expression only through the total phase factor $\Phi$. To get the
second equality, we first used the identities
$\sum_{\epsilon,\epsilon' = \pm} = \sum_{\epsilon,
\tilde{\epsilon}=\pm}$ ($\tilde{\epsilon} \equiv
\epsilon\epsilon'$) and then used
$e^{i\epsilon\omega_0(t-t_1)-i\epsilon'\omega_0t_2} =
e^{i\epsilon\omega_0\{t-t_1-\epsilon\epsilon' t_2\}}$.

\begin{figure}[t]
    \psfrag{t}{\small{$t$}}
    \psfrag{t1}{\small{$t_1$}}
    \psfrag{t2}{\small{$t_2$}}
    \psfrag{R1}{$R_{1}$}
    \psfrag{R2}{$R_{3}$}
    \psfrag{R3}{$R_{2}$}
    \includegraphics[width=0.25\textwidth]{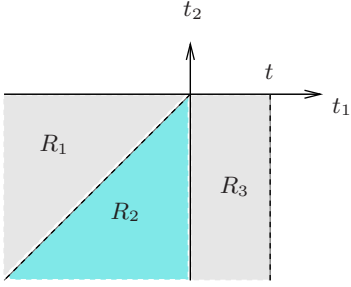}
    \caption{Subdomains of integration in the virtual time $(t_1,t_2)$ space as it is defined in Eq.~\eqref{eq:domains}. Only the virtual times $(t_1, t_2)$ that lie
    in the three shaded subdomains $R_{1}$, $R_{2}$, and $R_3$  give non-zero contribution to the cross current correlation Eq.~\eqref{eq:S-second} due to causality which is encoded in the Keldysh formalism.}
    \label{fig:map}
    \end{figure}
Now evaluating the summation over $\eta$'s:
    \begin{equation}
    \sum_{\eta_1,\eta_2=\pm1}\eta_1\eta_2\;e^{i\Phi^{\eta,\eta',\eta_1,\eta_2}_{\tilde{\epsilon}} (t,0,t_1,t_2)}
    \label{eq:sum-Phi}
    \end{equation}
for different ranges of $(t_1, t_2)$,allows us to extract the
statistical angle $\theta$ dependence of the cross correlation. To
do so, we begin by noting that the expression for $\Phi$
Eq.~\eqref{eq:Phi} takes a constant value for  different domains
and hence the sum over $\eta_1,\eta_2=\pm1$ in the brackets of
Eq.~\eqref{eq:I1I2-Phi} can be independently evaluated prior to
evaluating the integral if the virtual time $(t_1,t_2)$ space is
split into appropriate sub-domains.  Consequently, we make the
following observations:

 \begin{enumerate}
    \item We first note that $\Phi^{\eta,\eta',\eta_1,\eta_2}_{\tilde{\epsilon}} (t,0,t_1,t_2)$ is independent of $\eta_2$ for $t_2>0$.  Since $\sum_{\eta_2=\pm1}\eta_2=0$, the expression of Eq.~\eqref{eq:sum-Phi} vanishes identically for $t_2>0$. Hence the only non-zero contribution to the integral comes from $t_2<0$.

    \item A similar consideration for the summation over $\eta_1$ for $t_1>t$  allows us to
    limit the integration range to $t_1<t$ since
    $\Phi^{\eta,\eta',\eta_1,\eta_2}_{\tilde{\epsilon}} (t,0,t_1\!>\!t,t_2\!<\!0)$ is independent of $\eta_1$.

    \item The above analysis reflects the fact that
    only virtual times prior to the times of tunneling events can affect the
    correlation between tunneling events due to causality (thus acting as a check
     that our implementation of the Keldysh formalism for Klein factors is faithful in
     respecting causality).

    \item Now we are left only with the region $\{(t_1,t_2)|\; t_1\!<\!t,\, t_2\!<\!0\}$ which is shaded in the Fig.~\ref{fig:map}. In this region, the phase of Eq.~\eqref{eq:Phi} simplifies to
    \begin{equation}
    \Phi^{\eta,\eta',\eta_1,\eta_2}_{\tilde{\epsilon}}\! =\! \theta\left[\eta_1\left\{1\!+\!\tilde{\epsilon}\left(\frac{\sgn(t_1\!-\!t_2)\!-\!\sgn(t_1)}{2}\right)\!\right\}\!+\!\eta_2\right]
    \end{equation}
 where we used the fact that we are interested in the case $t>0$.  From this expression, which is  {\em independent of $\eta$ and $\eta'$}, we can identify three domains that contribute to the integral of Eq.~\eqref{eq:I1I2-Phi}:
    \begin{equation}
    \begin{split}
    &R_{1}\equiv\{(t_1,t_2)|\; t_1\!<\!t_2\!<\!0\}, \\
    &R_{2}\equiv\{(t_1,t_2)|\; t_2\!<\!t_1\!<\!0\}, \\
    &R_3\equiv\{(t_1,t_2)|\;t_2\!<\!0\!<\!t_1\!<\!t\}\
    \end{split}
    \label{eq:domains}
    \end{equation}
     as we depicted in Fig.~\ref{fig:map}.

    \item $\Phi^{\eta,\eta_1,\eta_2}_{\tilde{\epsilon}}(t,0,t_1,t_2)$ can now be considered as a function that depends on the domain $R_\zeta$ ($\zeta~=~1,2,3$) which can be evaluated to give
    \begin{align}
    &\Phi^{\eta,\eta',\eta_1,\eta_2}_{\tilde{\epsilon}}[R_{1}] = \theta(\eta_1+\eta_2),\label{eq:Phi12}\\
    &\Phi^{\eta,\eta',\eta_1,\eta_2}_{\tilde{\epsilon}}[R_{2}]=\theta\{\eta_1(1+\tilde{\epsilon})+\eta_2\},\label{eq:Phi21}\\
    &\Phi^{\eta,\eta',\eta_1,\eta_2}_{\tilde{\epsilon}}[R_{3}]=\theta(\eta_1+\eta_2). \label{eq:Phi0}
    \end{align}
    It is worth noting that $\Phi_{\tilde{\epsilon}}$ has an explicit dependence upon $\tilde{\epsilon}$ only in the domain $R_{2}$. However, since $\Phi^{\eta,\eta_1,\eta_2}_{\tilde{\epsilon}}[R_{2}]$ becomes independent of $\eta_1$ for $\tilde{\epsilon}=-1$, the sum Eq.~\eqref{eq:sum-Phi} will once again vanish. Hence we find that only the case $\tilde{\epsilon} = +1$, where the calculated correlation is between the tunneling processes with same orientations  (case $S$ of Fig.~\ref{fig:SO}), contributes to the part of the integral Eq.~\eqref{eq:I1I2-Phi} coming from the domain $R_{2}$. Later we will discuss the implication of this observation in more detail.

    \item Finally we can evaluate the phase factor summation over contour branch indices Eq.~\eqref{eq:sum-Phi} for each regions using Eqs.~(\ref{eq:Phi12}-\ref{eq:Phi0}) as the following:
    \begin{widetext}
    \begin{eqnarray}
    &{\text For }\; R_{1}:\; \sum_{\eta_1\eta_2=\pm1}\!\!\eta_1\eta_2e^{i\Phi^{\eta,\eta',\eta_1,\eta_2}_{\tilde{\epsilon}}[R_{1}]} &=
        \sum_{\eta_1,\eta_2=\pm1}\!\!\eta_1\eta_2e^{i\theta(\eta_1\!+\!\eta_2)} = -4\sin^2\theta
        \label{eq:phaseR1}\\
    &{\text For }\; R_{2}:\; \sum_{\eta_1\eta_2=\pm1}\!\!\eta_1\eta_2e^{i\Phi^{\eta,\eta',\eta_1,\eta_2}_{\tilde{\epsilon}}[R_{2}]}& =
        \sum_{\eta_1,\eta_2=\pm1}\!\!\eta_1\eta_2e^{i\theta(\eta_1(1+\tilde{\epsilon})+\eta_2)}
         = \left\{\begin{split}
            &-4\sin^2\theta(2\cos\theta)\;\; \text{for } \tilde{\epsilon}\!=\!+1 \\
            &0\;\; \mathrm{for }\;  \tilde{\epsilon}\!=\!-1
            \end{split} \right\}\label{eq:phaseR2}\\
    &{\text For }\; R_3:\;  \sum_{\eta_1\eta_2=\pm1}\!\!\eta_1\eta_2e^{i\Phi^{\eta,\eta',\eta_1,\eta_2}_{\tilde{\epsilon}}[R_3]} &=
        \sum_{\eta_1,\eta_2=\pm1}\!\!\eta_1\eta_2e^{i\theta(\eta_1\!+\!\eta_2)} = -4\sin^2\theta.\label{eq:phaseR3}
    \end{eqnarray}
    \end{widetext}
\end{enumerate}

At last, we can utilize the observations we made above to rewrite
the four-point contribution to the noise $S(t)$ spelled out in
Eq.~\eqref{eq:I1I2-Phi} in the following form
    \begin{multline}
    \langle I_1(t)I_2(0)\rangle^{(2)}\\
     =-8\sin^2\theta\, {e^*}^2|\Gamma_1\Gamma_2|^2\sum_{\tilde{\epsilon}=\pm}
     \int_{R_1, R_3}\!\!dt_1dt_2\tilde{\epsilon}\,\cos\omega_0(t-t_1-\tilde{\epsilon}t_2)\\
    \times\left(C(t-t_1)C(t_2)\right)^2\left[\frac{C(t_1)C(t-t_2)}{C(t_1-t_2)C(t)}\right]^{\tilde{\epsilon}}\\
    -16\sin^2\theta\cos\theta\,{e^*}^2|\Gamma_1\Gamma_2|^2\int_{R_2}dt_1dt_2
    \cos\omega_0(t-t_1-t_2)\\
    \times\left(C(t-t_1)C(t_2)\right)^2\left[\frac{C(t_1)C(t-t_2)}{C(t_1-t_2)C(t)}\right]
    \label{eq:I1I2-sumR}
    \end{multline}
In Eq.~\eqref{eq:I1I2-sumR} the statistical angle dependence has
been extracted out and we found an additional factor of
$2\cos\theta$ in the contribution from the domain $R_{2}$ which
does not occur in the contributions from domains $R_1$ and $R_3$.
We note that this contribution rises only from the case $S$ of
Fig.~\ref{fig:SO} (tunneling processes in the same orientation,
i.e. $\tilde{\epsilon}\!=\!+\!1$) as it is shown explicitly in the
second equality of Eq.~\eqref{eq:I1I2-sumR} by the value of
$\tilde{\epsilon}$ being set to $+1$ in the contribution (the
second term).

To complete our analysis, we now turn to the product of two point
functions (the second term of Eq.~\eqref{eq:St}). From
Eq.~\eqref{eq:I1}, the second term of Eq.~\eqref{eq:St} to order
$O(\Gamma_1^2\Gamma_2^2)$  becomes
    \begin{widetext}
    \begin{equation}
    \begin{split}
    &\langle I_1(t)\rangle\langle I_2(0)\rangle^{(2)}\\
    &\quad= -4\sin^2\theta\,{e^*}^2|\Gamma_1\Gamma_2|^2\sum_{\epsilon,\epsilon'=\pm}\epsilon\epsilon'\int dt_1dt_2\; e^{i\epsilon\omega_0(t-t_1)-i\epsilon'\omega_0(t_2)}
    C(t-t_1)^2C(t_2)^2.
    \end{split}
    \end{equation}
    \end{widetext}

Finally, putting all of the above together, the normalized noise
(cross current correlation) is found to be of the following form
    \begin{multline}
    S(\tilde{t};\,T/T_0,K)
    = \sum_{\zeta=1,3}\sum_{\tilde{\epsilon}=\pm} S^{\tilde{\epsilon}}_\zeta(\tilde{t};\,T/T_0,K) - 1 \\ +2\cos\theta\; S^+_2(\tilde{t};\,T/T_0,K),
    \label{eq:S-cos}
    \end{multline}
where $S^{\tilde{\epsilon}}_\zeta(t)$ for all values of $\tilde{\epsilon}=\pm$ and $\zeta=1,2,3$ is defined by an integral
over domain $R_\zeta$ for contributions with relative orientation
of tunneling $\tilde{\epsilon}$ normalized by the product of the
average tunneling current
    \begin{widetext}
    \begin{equation}
    S^{\tilde{\epsilon}}_\zeta(\tilde{t}; \, T/T_0, K) \equiv \frac{\D\tilde{\epsilon}\,\int_{R_\zeta}d\tilde{t}_1 d\tilde{t}_2\,\cos(\tilde{t}-\tilde{t}_1-\tilde{\epsilon}\tilde{t}_2)
    \left(C(\tilde{t}-\tilde{t}_1)C(\tilde{t}_2)\right)^2\left[\frac{C(\tilde{t}_1)C(\tilde{t}-\tilde{t}_2)}{C(\tilde{t}_1-t_2)C(\tilde{t})}\right]^{\tilde{\epsilon}}}
    {\D\sum_{\tilde{\epsilon}'=\pm1}\sum_{\zeta'=1,2,3}\int_{R_{\zeta'}}d\tilde{t}_1 d\tilde{t}_2\; \cos[\tilde{t}-\tilde{t}_1-\tilde{\epsilon}'\tilde{t}_2]C(\tilde{t}-\tilde{t}_1)^2C(\tilde{t}_2)^2}
    \label{eq:def-Szeta}
    \end{equation}
    \end{widetext}
where we defined the dimensionless time $\tilde{t}\equiv\omega_0t$
and also introduced a notation for the temperature equivalent to
the Josephson frequency to be $T_0\equiv\hslash\omega_0/k_B$ and
used the fact that $C(t; T,K)=C(\tilde{t}; T/T_0, K)$ from
Eq.~\eqref{eq:def-C}. Note that since we  chose to  study the
normalized noise, all cutoff dependence and tunneling amplitude
dependence in the noise is eliminated through normalization since
they enter as common factors for numerator and denominator of
Eq.~\eqref{eq:def-Szeta}.

 \subsection{Discussion}
 \label{subsec:AcosB}
  Now we can define two scaling functions $\EuScript{A}$ and $\EuScript{B}$ that depend on the scaling dimension $K$, the Josephson frequency $\omega_0=e^*V/\hslash$ (or $T_0$) and the temperature as
    \begin{align}
    &\EuScript{A}(\omega_0t;\,T/T_0,K)\equiv \sum_{\zeta=1,3}\sum_{\tilde{\epsilon}=\pm} S^{\tilde{\epsilon}}_\zeta(\omega_0t;\,T/T_0,K) - 1\nonumber\\
    &\EuScript{B}(\tilde{t};\,T/T_0,K)\equiv
    2S^+_2(\omega_0t;\,T/T_0,K).
    \label{eq:def-AB}
    \end{align}
We can thus rewrite Eq.~\eqref{eq:S-cos} as a sum of two scaling
functions,
 corresponding to a ``direct term'' and
an ``exchange term" which depends explicitly on  $\cos\theta$:
    \begin{multline}
    S\left(\omega_0t;\,\frac{T}{T_0},K\right)=\EuScript{A}\left(\omega_0t;\,\frac{T}{T_0},K\right) \\
    +
    \cos\theta\;\EuScript{B}\left(\omega_0t;\,\frac{T}{T_0},K\right),
    \label{eq:St-AB}
    \end{multline}
where $T_0\!=\!\hslash\omega_0/k_B$. Hence we were able to extract
the statistical angle dependence of the noise in the form of the
factor of $\cos\theta$ in the second term while both of the
scaling functions $\EuScript{A}$ and $\EuScript{B}$ {\em do not}
depend on the statistics. Eq.~\eqref{eq:St-AB} is the key result
of this paper. One can gain further insight by recognizing the
fact that the domain $R_1$ (which contributes to  $\EuScript{A}$)
and $R_2$ (which contributes to $\EuScript{B}$) are related to
each other via exchange of virtual times $t_1$ and $t_2$ (See
Fig.~\ref{fig:map}). In the following we discuss the physics
behind the factor $\cos\theta$ in Eq.~\eqref{eq:St-AB} with an eye
towards bridging the connection between {\em exclusion statistics}
and {\em exchange statistics} in our setup.

This expression displays a number of noteworthy properties:
\renewcommand{\labelenumi}{\alph{enumi}.}
\begin{enumerate}
\item This finite temperature result is
applicable to all Jain states (in contrast to recent zero
temperature results on Laughlin
states\cite{safi01,vishveshwara03}) and it is a universal scaling function to the lowest order in perturbation theory which is valid provided the tunneling current is small compared to
the Hall current. 

\item Fractional charge and statistics play fundamentally distinct roles,
each entering Eq.~\eqref{eq:St-AB} through different features:
fractional charge through $\omega_0=e^*V/\hslash$ and fractional
statistics through the $\cos\theta$ factor.
\item Given that  for  Laughlin states $\theta<\pi/2$ and
$\cos\theta>0$, the exchange term in Eq~\eqref{eq:St-AB} provides
largely positive (``bunching'', boson-like) contributions to the
noise whereas for non-Laughlin states with $\theta>\pi/2$, its
contribution is largely negative (``anti-bunching'',
fermion-like).
\item The fact that only case $S$, of Fig.~\ref{fig:hbt}, contributes to this factor
 can be viewed as a manifestation of a
generalized exclusion principle\cite{haldane91} since the virtual
processes in this case involves adding a qp to edge $0$ in the
presence of another. It is noteworthy that we arrived at an
observable consequence of a generalized {\em exclusion} principle
given that Eq.~\eqref{eq:St-AB}) was derived using anyonic
commutation rules prescribed by {\it braiding}
statistics~\cite{arovas84}.
\end{enumerate}

A few remarks are in order here. For the function
in Eq.~\eqref{eq:St-AB} to be observable, given that it is the
ratio between the cross-current correlations and the average
values of the currents involved, the former needs to be at least
comparable to the latter. As mentioned above, the form of the
function only takes into account the lowest order terms in
tunneling; when higher order terms become important, the function
no longer remains universal in that it exhibits an explicit
dependence on tunneling amplitudes. We assumed that tunneling from
edge $0$ occurred from a single point. In our calculations, we
could have in principle retained the separation scale '$a$' for
points from which qps tunnel into edges $1$ and $2$, respectively.
Then, $1/a$ would enter the problem as another energy scale.  The scaling function would depend on another
parameter $\omega_0a$ and
the cross-current correlations would decay with separation length. 

 As one might expect from the simple form of the
statistical angle dependence of Eq.~\eqref{eq:St-AB}, this
expression can be understood in terms of simple pictures in which
the setup simultaneously acts as an accessible stage for exchange
processes and a testbed for generalized exclusion.

To see that the form of Eq.~\eqref{eq:St-AB}
derives directly from {\em exchange} of qp's between edge $1$ and
edge $2$ via edge $0$, we first note that information regarding
anyonic exchange statistics and related time ordering of events is
carried as a phase factor in the qp propagators. Contour ordering
of events and associated time-dependent shuffling is completely
contained in this exchange statistics phase factor. The following
analysis of tunneling events and their effect on the phase factor
not only shows that the simple $\cos\theta$ factor in
Eq.~\eqref{eq:St-AB} naturally emerges from the events associated
with exchange statistics, but that it is also a manifestation of
exclusion statistics.

Given that causality constrains lowest order virtual processes to
occur at times $t_1$ and $t_2$ within domains $R_1$, $R_2$ and
$R_3$ of Fig.~\ref{fig:map}, a qp/qp from edge $1$ cannot exchange
with a qp/qh from edge $2$ when $t_1>0$ (domain $R_3$) since qp/qh
tunnels between edge $2$ and edge $0$ via virtual process at time
$t_2<0$ and back at  time $0$ before anything happens between $1$
and $0$. Hence it is clear as to why $S^{\tilde{\epsilon}}_3$
coming from domain $R_3$ contributes to $\EuScript{A}$ and an
uncorrelated two-point piece is subtracted out.

Now for domains $R_1$ and $R_2$, there is an overlap $|t_1-t_2|$
in the times during which qp/qh's from edge 1 and edge 2 stay in
edge $0$ and hence can experience statistics induced interference.
Between $R_1$ and $R_2$, the only pairs of time whose relative
orders change from one domain to the other is $t_1$ and $t_2$ and
hence $R_1$ and $R_2$ are related to each other via exchange of
entities involved in the virtual processes occurring at $t_1$ and
$t_2$. For both domains,   depending on the relative orientations
of the currents, the virtual tunneling processes leave two
distinct pairs of objects at edge $0$ as shown in
Fig.~\ref{fig:SO}.

In case $S$ ($\tilde{\epsilon}=+1$), where $I_1$ and $I_2$ have
the same orientation , the entities left behind are a pair of two
qp's. Hence the domain $R_2$ is related to the domain $R_1$ via
exchange of these two identical particles.  This exchange between
two qp's brings in an additional phase factor of
$e^{i\eta_1\theta}$ in Eq.~\eqref{eq:phaseR2} for the case
$\tilde{\epsilon}=+$ that adds (constructive interference) to the
phase factor $e^{i\eta_1\theta+i\eta_2\theta}$ that is common to
all three domains (See Eqs.~(\ref{eq:phaseR1}-\ref{eq:phaseR3}).)
and also to the uncorrelated piece.  Since the $\tilde{\epsilon}$
independent phase factor sum for domains $R_1$,  $-4\sin^2\theta$,
is the same as that of the uncorrelated piece that normalizes the
noise (and that for the domain $R_3$), it cancels upon
normalization while the phase factor sum from the case $S$ of
domain $R_2$ results in the factor of
$2\cos\theta=\sin2\theta/\sin\theta$. As a consequence
$S^{\tilde{\epsilon}}_1$ is a part of $\eA$ without statistical
angle dependence while $S^{+}_2$ enters $\EuScript{B}$. In case
$O$ ($\tilde{\epsilon}=-1$), on the other hand, tunneling currents
have opposite  orientations leaving one qp and one qh behind.
Exchange between qp and qh brings in an additional phase of
$e^{-i\eta_1\theta}$ in $R_2$ (See Eq.~\eqref{eq:phaseR2} for
$\tilde{\epsilon}=-$) that cancels with the common phase factor
$e^{i\eta_1\theta}$ (destructive interference) resulting in
$\sin(\theta-\theta) = 0$ for the phase factor sum. Hence there is
no $S^{-}_2$ contribution to $\EuScript{B}$ but only $S^{+}_2$
constitutes $\EuScript{B}$.

The generalized exclusion statistics is defined by counting the
change in the one particle Hilbert space dimensions $d_\alpha$ for
particles of species $\alpha$ as particles are added (keeping the
boundary conditions and the size of the system constant) in the
presence of other particles \cite{haldane91,vanelburg98}. The
defining quantity for this description of fractional statistics is
the {\em statistical interaction }$g_{\alpha\beta}$
\begin{equation}
\Delta d_a = -\sum_\beta g_{\alpha\beta}\Delta N_\beta,
\label{eq:exclusion}
\end{equation}
where $\{\Delta N_\beta\}$ is a set of allowed changes of the
particle numbers at fixed size and boundary conditions boundary
conditions. For bosons (without hardcore) $g_{\alpha\beta}=0$,
while the Pauli exclusion principle for fermions correspond to
$g_{\alpha\beta}=\delta_{\alpha\beta}$. Thus, while the presence
of bosons does not change the available number of states for an
additional boson, the presence of a fermion reduces the number of
available states by one. Once one understands the one particle
Hilbert space of the particle of interest, the statistical
interaction $g_{\alpha\beta}$ can be calculated through state
counting and for FQH qp's such program can be carried out using
conformal field theory (CFT) as shown by \textcite{vanelburg98}.

However, the connection between this definition and an
experimentally measurable quantity has not been clear so far.
Further, the connection between fractional statistics defined
through {\em exclusion} and anyonic statistics defined through
{\em exchange} has not been clearly established for the following
reason.  The anyonic exchange  for FQH qp's  based on braid groups
defined in Refs.~[\onlinecite{leinaas77,wilczek82} ] is achieved
through attaching Chern-Simons flux to hard-core
bosons\cite{zhang89} or spinless fermions\cite{lopez91}. However
how such flux attachments would affect Hilbert space dimensions is
not obvious.  For Laughlin qp's, one can use the knowledge of the
chiral Hilbert space from CFT\cite{vanelburg98} to find that the
two definitions coincide. Nonetheless, an equivalent understanding
for non-Lauhglin states are still lacking.

 In our setup, it is quite clear that only virtual processes occurring
 in domains $R_1$ and $R_2$ with $\tilde{\epsilon}=+1$ ( case $S$ in Fig.~\ref{fig:SO}) host the situation of a qp being added in the presence of the other. The difference between $R_1$ and $R_2$ is the following:
 In  $R_1$, a qp from edge $2$ tunnels in to edge $0$ at time $t_2$ and leaves
 at time $0$ with the qp from edge $1$ present through out. In $R_2$, however, a qp from edge $1$ tunnels in to the edge $0$ at time $t_1$ later ``pushing'' a qp to edge $2$ at time $t_2$.
Therefore virtual processes occurring in domain $R_2$ should be
affected by {\em statistical interaction}
 of the generalized exclusion principle and we can understand why $S^{+}_2$ alone consists the statistics dependent term
from this simple reasoning.

It is astonishingly consistent that the processes contributing to
noise in which the generalized exclusion principle should manifest
itself are the ones that are associated with the simple factor
$\cos\theta$ considering that the statistical angle is what {\em
defines} exchange (braiding) statistics.  The fact that our
results describe an experimentally measurable quantity makes it
even more interesting. Although the manner in which the factor can
be related to $g_{\alpha\beta}$ defined in Eq~\eqref{eq:exclusion}
needs further investigation, it is clear from our choice of setup
which allows access to the phase gain upon exchange while
satisfying conditions for observation of generalized exclusion
principle, that the connection exists. In the following sections,
we analyze the scaling functions $\EuScript{A}(t)$ and
$\EuScript{B}(t)$ after carrying out the integration over
respective domains numerically and show that the noise experiment
of the type proposed here can be used to detect and measure
fractional statistics.

 \section{The time dependent noise $S(t)$ analysis}
 \label{sec:St}
Since the definite double integral in the numerator of the
expression for $S^{\tilde{\epsilon}}_\zeta$
Eq.~\eqref{eq:def-Szeta} does not grant an analytic expression in
closed form, we evaluate $S^{\tilde{\epsilon}}_\zeta$ numerically.
The integrable singularity in the vicinity of the domain boundary
is regulated through a short time cutoff $\tau_0$. A natural
choice for this cutoff that is intrinsic to QH system would be
 the inverse cyclotron frequency $\omega_c^{-1}$, with $\omega_c\hslash\sim 1\text {m}eV$ for the magnetic field of order $10 T$  required for the FQH.
As we will discuss later, given that $\omega_c$ is such a high
energy scale when compared to other important energy scales of the
setup, namely the Josephson frequency $\omega_0\hslash\equiv e^*V$
and the temperature $k_BT$, as ought to be the case, all
interesting features of the noise turn out to be cutoff
independent.

Before we discuss the results of our numerical integration, we can
gain some insight by looking into the asymptotic behavior of
$S^{\tilde{\epsilon}}_\zeta(\omega_0 t)$ in the short time limit
${k_BTt,\, \omega_0t}\ll 1$ and in the long time limit
${k_BTt,\,\omega_0t}\gg1$. In the short time limit, the numerator
of Eq.~\eqref{eq:def-Szeta} becomes
\begin{multline}
S^{\tilde{\epsilon}}_\zeta(t)\propto \tilde{\epsilon}\,
|k_BTt|^{\tilde{\epsilon}K}
\int_{R_\zeta}\!\!d\tilde{t}_1d\tilde{t}_2
\cos(\tilde{t}_1\!+\!\tilde{\epsilon}\tilde{t}_2)\\
(C(\tilde{t}_1)C(\tilde{t}_2))^2\left|\frac{C(\tilde{t}_1)C(\tilde{t}_2)}{C(\tilde{t}_1\!-\!\tilde{t}_2)}\right|^{\tilde{\epsilon}}
\label{eq:St-short-t}
\end{multline}
where we used
\begin{equation}
\lim_{t\rightarrow 0} C(t) \propto |\pi k_BT t|^{-K}.
\label{eq:C-short-t}
\end{equation}
Since the domain size of $R_3$ is proportional to $t$, the
integral in Eq.~\eqref{eq:St-short-t} is negligible for $\zeta=3$
and hence we can ignore $S^{\tilde{\epsilon}}_3$'s contribution to
$\eA$
 as $t\rightarrow 0 $.
However since the integral in Eq.~\eqref{eq:St-short-t} is
independent of $t$ for domains $R_1$ and $R_2$,  $S^{-}_1(t)$
diverges as $(k_BTt)^{-K}$ for $\tilde{\epsilon}=-$ as
$t\rightarrow 0$ and both $S^{+}_1(t)$ and  $S^{+}_2$ vanish as
$(k_BTt)^{K}$. As a result, $\EuScript{A}(t)$ defined in
Eq.~\eqref{eq:def-AB} is dominated by $S^{-}_1$ in the short time
limit and we find
\begin{equation}
|\EuScript{A}(t)|\sim |k_BTt|^{-K}(1+\cdots). \label{eq:A-short-t}
\end{equation}
On the other hand,
\begin{equation}
|\EuScript{B}(t)|=2|S^{+}_2(t)|\sim |k_BTt|^K.
\label{eq:B-short-t}
\end{equation}

In the long time limit, the numerator of Eq.~\eqref{eq:def-Szeta}
can be approximated for $\zeta=1,2$  ($S^{\tilde{\epsilon}}_1$ and
$S^{+}_2$) by
\begin{multline}
S^{\tilde{\epsilon}}_\zeta(t)\propto \tilde{\epsilon}\,
e^{-2K\pi|k_BTt|} \int_{R_\zeta}\!\!d\tilde{t}_1d\tilde{t}_2
\cos(t\!-\!\tilde{t}_1\!-\!\tilde{\epsilon}\tilde{t}_2)\\
C(\tilde{t}_2)^2\left|\frac{C(\tilde{t}_1)}{C(\tilde{t}_1\!-\!\tilde{t}_2)}\right|^{\tilde{\epsilon}}
\label{eq:St-long-t}
\end{multline}
which vanishes as $e^{-2K\pi|k_BTt|}$ multiplied by an oscillating
factor.
 Also, since the size of the domain $R_3$ grows with $t$, $S^{+}_3$ and $S^{-}_3$ become finite and
 we find
 \begin{equation}
 \lim_{t\rightarrow\infty} [S^{+}_3(t)+S^{-}_3(t)] - 1 =0.
 \end{equation}
 Therefore one can expect both $\EuScript{A}(t)$ and $\EuScript{B}(t)$ to vanish exponentially
 in the long time limit due to thermal fluctuations.

\begin{figure}[htb]
\subfigure[High temperatures.] { \psfrag{w}{\small$\omega_0t$}
\psfrag{S}{$\EuScript{A}(\omega_0t)$, $\EuScript{B}(\omega_0t)$}
\psfrag{0.25}{\footnotesize$0.25$}
\psfrag{-0.25}{\footnotesize$-0.25$}
\psfrag{-0.5}{\footnotesize$-0.5$} \psfrag{2}{\footnotesize$2$}
\psfrag{4}{\footnotesize$4$} \psfrag{6}{\footnotesize$6$}
\includegraphics[width=0.35\textwidth]{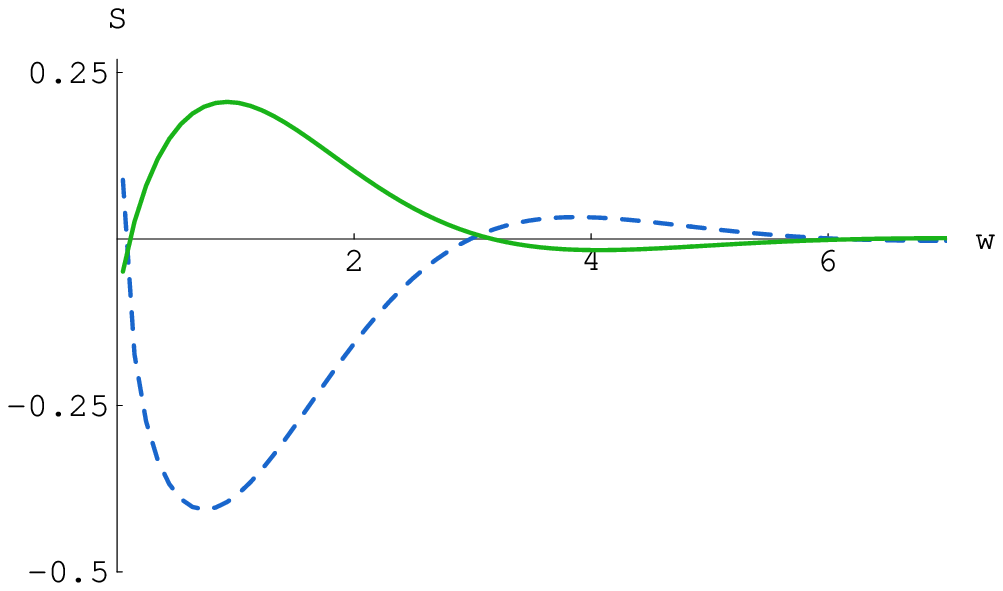}
} \subfigure[Low temperatures.] { \psfrag{w}{\small$\omega_0t$}
\psfrag{S}{$\EuScript{A}(\omega_0t)$, $\EuScript{B}(\omega_0t)$}
\psfrag{1}{\footnotesize$1$} \psfrag{-1}{\footnotesize$-1$}
\psfrag{-2}{\footnotesize$-2$} \psfrag{10}{\footnotesize$10$}
\psfrag{20}{\footnotesize$20$} \psfrag{30}{\footnotesize$30$}
\psfrag{40}{\footnotesize$40$}
\includegraphics[width=0.35\textwidth]{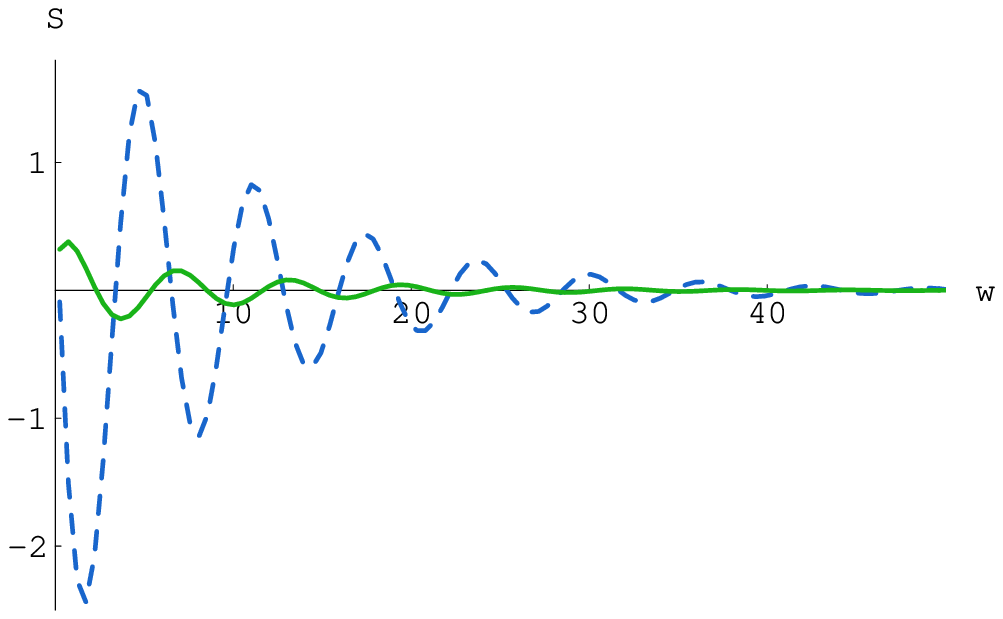}
} \caption{\label{fig:St}The scaling functions
$\EuScript{A}(\omega_0t)$ (blue dashed line) and
$\EuScript{B}(\omega_0t)$ (green solid line) for $\nu=1/5$ as a
function of dimensionless time $\omega_0 t$ at two different
temperatures: (a) $T/T_0=2$  and (b)$T/T_0=0.25$ where the
reference scale is set by the Josephson frequency to be $T_0\equiv
\frac{\omega_0\hslash}{\pi k_B}$.
 $\EuScript{A}(\omega_0t)$
is dominantly negative at short times due to the divergent
contribution coming from counter-current correlations while
$\EuScript{B}(\omega_0t)$ coming only from correlations of
currents in the same direction is dominantly positive at short
times. $\EuScript{A}(\omega_0t)$ and $\EuScript{B}(\omega_0t)$ for
other filling factors show similar forms up to an overall scale
set by the associated scaling dimension $K$. }
\end{figure}

Depending on the relative magnitude of the two energy scales of
the problem, namely the Josephson frequency $\omega_0\hslash=e^*V$
and temperature $k_BT$, there may or may not be an intermediate
range of times over which both $\EuScript{A}(t)$ and
$\EuScript{B}(t)$ show oscillatory behavior. Figure~\ref{fig:St}
shows the typical time dependence of $\EuScript{A}(\omega_0t)$ and
$\EuScript{B}(\omega_0t)$ as a function of dimensionless time
$\omega_0t\equiv\tilde{t}$ at high temperatures
$\omega_0\hslash\sim k_BT$ and low temperatures
$\omega_0\hslash\gg k_BT$.

We can confirm from Fig.~\ref{fig:St} (a) that when the
characteristic  time for the exponential decay in the long time
limit  $(2K\pi k_B T)^{-1}$ (See Eq.~\eqref{eq:St-long-t}) is of
the same order or less than the oscillation period
$\omega_0^{-1}$, no oscillations are seen. At the same time, the
short time behavior which sets in only for $t\ll (k_BT)^{-1}$ is a
less prominent feature of the plot. At moderate time scales
however,  $\EuScript{A}(t)$ and $\EuScript{B}(t)$ are of
comparable magnitude hence the $\cos\theta$ factor in the
``exchange term'' of Eq.~\eqref{eq:St-AB} can significantly shift
the total noise positively or negatively depending on whether
$\theta<\pi/2$ or $\theta>\pi/2$. Hence it is important to
understand the intermediate time scale and it can be done so
numerically.

For temperatures $T\ll T_0=\omega_0\hslash/k_B$,  both
$\EuScript{A}(t)$  and $\EuScript{B}(t)$ in Fig.~\ref{fig:St} (b)
demonstrate definite oscillatory behavior characterized by the
Josephson frequency $\omega_0$ with an exponentially decaying
envelope $e^{-K\pi k_BTt}$. Although the magnitude of
$\EuScript{B}(t)$ is much smaller than that of $\EuScript{A}(t)$
due to the dominant short-time divergent contributions to
$\EuScript{A}(t)$ from $S^{-}_1$, one can expect the Fourier
spectrum of $\widetilde{\EuScript{A}}(\omega)$ and
$\widetilde{\EuScript{B}}(\omega)$ to show singular behavior in
the vicinity of $\omega=\omega_0$. Such features of the Fourier
spectrum can be used in experiments to locate the Josephson
frequency which is determined by the fractional charge of the FQH
qp. In the next section, we discuss the Fourier spectra
$\widetilde{\EuScript{A}}(\omega)$ and
$\widetilde{\EuScript{B}}(\omega)$ with a particular interest in
extracting signatures of fractional statistics and fractional
charge.

\section{Frequency spectrum $\widetilde{S}(\omega)$}
\label{sec:Sw}
As with the time dependent noise, it can be seen from
Eq.~\eqref{eq:S-cos} that the frequency spectrum of the normalized
noise
 also gets contributions $\widetilde{S}^{\tilde{\epsilon}}_{\zeta}$ from different domains $R_\zeta$ as the following:
\begin{equation}
\widetilde{S}(\omega)=\int_{-\infty}^\infty dt e^{i\omega t}
S(t)\equiv \sum_{\zeta=1,3}\sum_{\tilde{\epsilon}=\pm}
\widetilde{S}^{\tilde{\epsilon}}_\zeta(\omega)
+2\cos\theta\,\widetilde{S}^{+}_2(\omega).
\end{equation}
Noting the fact that the oscillatory time dependence of all
$S^{\tilde{\epsilon}}_\zeta(t)$ comes from the factor
$\cos[\omega_0(t-t_1-\tilde{\epsilon}t_2)]$ in
Eq.~\eqref{eq:def-Szeta} ($\tilde{t}\equiv\omega_0t$), we can use
the identity
\begin{multline}
\cos[\omega_0(t-t_1-\tilde{\epsilon}t_2)] =
\cos(\omega_0t)\cos[\omega_0(t_1+\tilde{\epsilon}t_2)]
\\+\sin(\omega_0t)\sin[\omega_0(t_1+\tilde{\epsilon}t_2)]
\label{eq:trig}
\end{multline}
 to factor out oscillatory factors from  each $S^{\tilde{\epsilon}}_{\zeta}(t)$ as the following:
 \begin{equation}
 S^{\tilde{\epsilon}}_{\zeta}(t)
 \propto\left[\cos(\omega_0t)\eF^{\tilde{\epsilon}}_{\zeta}(t)
 +\sin(\omega_0t)\eH^{\tilde{\epsilon}}_{\zeta}(t)\right],
 \label{eq:S-FH}
 \end{equation}
where the  decayin  envelope functions
$\eF^{\tilde{\epsilon}}_{\zeta}(t)$ and
$\eH^{\tilde{\epsilon}}_{\zeta}(t)$ are defined by
\begin{widetext}
\begin{align}
\eF^{\tilde{\epsilon}}_{\zeta}(t)&\equiv
\tilde{\epsilon}\int_{R_\zeta}\!dt_1dt_2\,\cos[\omega_0(t_1+\tilde{\epsilon}t_2)]
\left(C(\tilde{t}-\tilde{t}_1)C(\tilde{t}_2)\right)^2\left[\frac{C(\tilde{t}_1)C(\tilde{t}-\tilde{t}_2)}{C(\tilde{t}_1-t_2)C(\tilde{t})}\right]^{\tilde{\epsilon}}\\
 \eH^{\tilde{\epsilon}}_{\zeta}(t)&\equiv \tilde{\epsilon}\int_{R_\zeta}\!dt_1dt_2\,\sin[\omega_0(t_1+\tilde{\epsilon}t_2)]
 \left(C(\tilde{t}-\tilde{t}_1)C(\tilde{t}_2)\right)^2\left[\frac{C(\tilde{t}_1)C(\tilde{t}-\tilde{t}_2)}{C(\tilde{t}_1-t_2)C(\tilde{t})}\right]^{\tilde{\epsilon}}
\label{eq:def-FH}
\end{align}
\end{widetext}
from Eq.~\eqref{eq:trig} and Eq.~\eqref{eq:def-Szeta}.

It is clear from Eq.~\eqref{eq:S-FH} that
$\widetilde{S}(\omega)^{\tilde{\epsilon}}_\zeta$ is a convolution
of an oscillating part and the envelopes which we may simply
compute. Thus we have
\begin{multline}
\widetilde{S}(\omega)^{\tilde{\epsilon}}_\zeta\propto
\frac{1}{2}\left[\widetilde{\eF}^{\tilde{\epsilon}}_\zeta(\omega+\omega_0)+\widetilde{\eF}^{\tilde{\epsilon}}_\zeta(\omega-\omega_0)\right]\\
+\frac{1}{2i}\left[\widetilde{\eH}^{\tilde{\epsilon}}_\zeta(\omega+\omega_0)+\widetilde{\eH}^{\tilde{\epsilon}}_\zeta(\omega-\omega_0)\right],
\end{multline}
with
\begin{align}
&\widetilde{\eF}^{\tilde{\epsilon}}_\zeta(\omega)
=\int_{-\infty}^\infty\!\! dt\,
\eF^{\tilde{\epsilon}}_\zeta(t)\,e^{i\omega t} =
2\int_0^\infty\!\! dt\, \cos\omega
t\;\eF^{\tilde{\epsilon}}_\zeta(t)\nonumber\\
&\widetilde{\eH}^{\tilde{\epsilon}}_\zeta(\omega)
=\int_{-\infty}^\infty\!\! dt\,
\eH^{\tilde{\epsilon}}_\zeta(t)\,e^{i\omega t} =
2i\int_0^\infty\!\! dt\, \sin\omega
t\;\eH^{\tilde{\epsilon}}_\zeta(t), \label{eq:tilde-FH}
\end{align}
where we used the fact that $ \eF^{\tilde{\epsilon}}_\zeta(t)$ is
even in $t$ while  $\eH^{\tilde{\epsilon}}_\zeta(t)$ is odd in
$t$. Hence we can understand the frequency spectrum by
investigating frequency spectra
$\widetilde{\eF}^{\tilde{\epsilon}}_\zeta(\omega)$,
$\widetilde{\eH}^{\tilde{\epsilon}}_\zeta(\omega)$ and superposing
them with their centers shifted by $\pm\omega_0$. One important
observation to be made in Eq.~\eqref{eq:tilde-FH} is the fact that
$\widetilde{\eF}^{\tilde{\epsilon}}_\zeta(\omega)$ is an {\em
even} function of $\omega$ while
$\widetilde{\eH}^{\tilde{\epsilon}}_\zeta(\omega)$ is an {\em odd}
function of $\omega$.
As we discussed in the previous section (See
Eq.~\eqref{eq:St-long-t}), $S^{\tilde{\epsilon}}_\zeta(t)$ decays
exponentially as $e^{-2K\pi k_BTt}$ in the long time limit  $(K\pi
k_BTt)\gg 1$ and hence
\begin{equation}
F^{\tilde{\epsilon}}_\zeta(t)\sim
\eH^{\tilde{\epsilon}}_\zeta(t)\rightarrow e^{-2K\pi k_BTt}
\end{equation}
in the  long time limit as well. As a result, the form of
$\widetilde{\eF}^{\tilde{\epsilon}}_\zeta(\omega)$ is a broadened
peak at $\omega=0$ with the width of the peak $2K\pi k_BT$
determined by the temperature and the scaling dimension of FQH qp.
Similarly, $\widetilde{\eH}^{\tilde{\epsilon}}_\zeta(\omega)$
takes the form of a derivative of a peak with the same width. The
characteristic behavior of
$\widetilde{\eF}^{\tilde{\epsilon}}_\zeta(\omega)$ and
$\widetilde{\eH}^{\tilde{\epsilon}}_\zeta(\omega)$ is plotted in
the Figure~\ref{fig:FH}.

 From the definition of  the ``direct'' and ``exchange'' terms in Eq.~\eqref{eq:def-AB}, we can express
 $\widetilde{\eA}(\omega)$ and $\widetilde{\eB}(\omega)$ in terms of $\widetilde{\eF}^{\tilde{\epsilon}}_\zeta(\omega)$ and $\widetilde{\eH}^{\tilde{\epsilon}}_\zeta(\omega)$
 as
    \begin{widetext}
    \begin{align}
    \widetilde{\eA}(\omega) & = \sum_{\zeta=1,3}\sum_{\tilde{\epsilon}=\pm}
    \frac{1}{2} \left[ \widetilde{\eF}^{\tilde{\epsilon}}_\zeta(\omega_0+\omega;\,T,K)
    +\widetilde{\eF}^{\tilde{\epsilon}}_\zeta(\omega_0-\omega;\,T,K) \right]\nonumber\\
    &\quad+\sum_{\zeta=1,3}\sum_{\tilde{\epsilon}=\pm} \frac{1}{2i}\left[ \widetilde{ \eH}^{\tilde{\epsilon}}_\zeta(\omega_0+\omega;\,T,K)
        -\widetilde{ \eH}^{\tilde{\epsilon}}_\zeta(\omega_0-\omega;\,T,K) \right]   \\
    \widetilde{\eB}(\omega) & = \frac{1}{2} \left[ \widetilde{\eF}^{+}_2(\omega_0+\omega;\,T,K)
    +\widetilde{\eF}^{+}_2(\omega_0-\omega;\,T,K) \right]   \nonumber\\
    &\quad+\frac{1}{2i}\left[ \widetilde{ \eH}^{+}_2(\omega_0+\omega;\,T,K)
        -\widetilde{ \eH}^{+}_2(\omega_0-\omega;\,T,K) \right].
    \end{align}
    \end{widetext}
In the rest of this section, we use the properties of
$\widetilde{\eF}^{\tilde{\epsilon}}_\zeta(\omega)$ and
$\widetilde{\eH}^{\tilde{\epsilon}}_\zeta(\omega)$ to understand
and discuss $\widetilde{\eA}(\omega)$ and
$\widetilde{\eB}(\omega)$   in two frequency regimes of interest:
low frequency  $\omega\ll\omega_0$ and near Josephson frequency
$\omega\sim\omega_0$.

\begin{figure}[h]
\psfrag{S}{\footnotesize$\widetilde \eA_0$, $\widetilde \eB_0$ }
\psfrag{T}{\footnotesize$T/T_0$} \psfrag{1}{\tiny$1$}
\psfrag{2}{\tiny$2$} \psfrag{3}{\tiny$3$} \psfrag{4}{\tiny$4$}
\psfrag{-4}{\tiny$-4$} \psfrag{-8}{\tiny$-8$}
\includegraphics[width=0.32\textwidth]{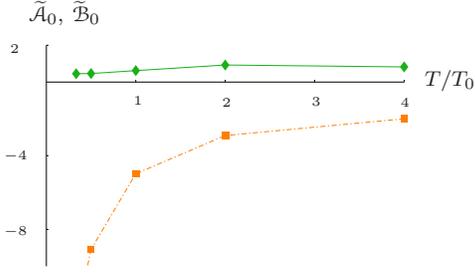}
\caption{\label{fig:Sw0} A typical temperature dependence of zero
frequency values of the direct term $\widetilde \eA(\omega\!=\!0)$
(orange dashed line) and the exchange term $\widetilde
\eB(\omega\!=\!0)$ (green solid line) taken from $\nu=2/5$ case.
The exchange contribution $\widetilde \eB(\omega\!=\!0)$ is
comparable to the direct term in the  temperature range $2\lesssim
T/T_0\lesssim4$ and hence the fractional statistics has a marked
effect on the total noise at small frequencies in the same
temperature range. The qualitative behavior is universal to all
fractions.  }
\end{figure}

   For  $\omega\ll\omega_0$, both $\widetilde{\eA}(\omega)$ and $\widetilde{\eB}(\omega)$ collect tail ends of shifted $\widetilde{\eF}^{\tilde{\epsilon}}_\zeta(\omega)$ and $\widetilde{\eH}^{\tilde{\epsilon}}_\zeta(\omega)$. Naturally, they
 show white noise behavior and hence we focus on zero
 frequency values $\widetilde \eA(\omega\!=\!0)\equiv\widetilde \eA_0$ and $\widetilde \eB(\omega\!=\!0)\equiv\widetilde \eB_0$ which depend only on the temperature(See Fig.~\ref{fig:Sw0}).
The direct term $\widetilde \eA_0$ is negative due to its dominant
contributions coming from $\tilde{\epsilon}=-$ processes in domain
$R_1$ with the opposite orientation of the  currents as
illustrated in case $O$ of Fig.~\ref{fig:SO}. On the other hand,
$\widetilde \eB_0$ is positive as it only involves case $S
(\tilde{\epsilon}=+)$ from domain $R_2$.  As a result,
Eq.~\eqref{eq:St-AB} translates into
\begin{equation}
\widetilde S(\omega\!=\!0)=-|\widetilde
\eA_0|+\cos\theta|\widetilde \eB_0|
\end{equation}
 for the frequency spectrum of the noise at zero frequency.
At low temperatures ($T\ll T_0$, $T_0\equiv
\hslash\omega_0/k_B\sim80{\text m}K$ for $V\!=\!40\mu V$),
$|\widetilde \eA_0|$ dominates the spectrum since contributions
from $\widetilde{\eH}^{-}_1(\omega_0)$ and
$\widetilde{\eF}^{-}_1(\omega_0)$  both diverges in the
$T\rightarrow0$ limit displaying $T^{-K}$ power law divergence.
However at higher temperatures, both
$\widetilde{\eF}^{\tilde{\epsilon}}_\zeta(\omega)$  and
$\widetilde{\eH}^{\tilde{\epsilon}}_\zeta(\omega)$ are more like
Lorenzian  for all $\tilde{\epsilon}$ and $\zeta$. Hence for
temperatures  of order $T_0$, $\widetilde{\eB}_0$ is comparable to
$\widetilde{\eA}_0$ as we display in Fig.~\ref{fig:Sw0}.  (At even
higher temperatures, both correlations are washed out by thermal
fluctuations.) Consequently the statistics dependent exchange
contribution, $\cos\theta\,|\widetilde \eB_0|$, will noticeably
affect the experimental measurement of the total noise at small
frequencies at such temperatures.
 It is rather astonishing that the statistical angle $\theta$,
 which was originally quantum mechanically defined in a highly
 theoretical basis, can be measured in this reasonably accessible frequency range at finite temperature.

Near the Josephson frequency,  $\widetilde \eA$ and $\widetilde
\eB$ develop qualitatively different sharp features at
temperatures $T\ll T_0/K$: $\widetilde \eA(\omega)$
nearly crosses $\omega$ axis steeply while $\widetilde
\eB(\omega)$ develops a peak  of width $2K
k_BT/\hslash$. This can be understood by analyzing which
terms among $\widetilde{\eF}^{\tilde{\epsilon}}_\zeta$ and
$\widetilde{\eH}^{\tilde{\epsilon}}_\zeta$ dominate each of
$\widetilde \eA(\omega)$ and $\widetilde
\eB(\omega)$. As for $\widetilde\eA$, its dominant
contribution is from domain $R_1$ with $\tilde{\epsilon}=-$.
Further,
$i\widetilde{\eH}^{-}_1(\omega)$ is strongly dominant over $\widetilde{\eF}^{-}_1(\omega)$  in magnitude as can be reasoned as below.

Recalling from Eq.~\eqref{eq:def-FH} that
$\widetilde{\eF}^{-}_1(\omega)$ and
$i\widetilde{\eH}^{-}_1(\omega)$ have
$\cos\left(\omega_0(t_1-t_2)\right)\left|\sinh(\pi
k_BT(t_1-t_2))\right|^{-K}$ and
$\sin\left(\omega_0(t_1-t_2)\right)\left|\sinh(\pi k_B
T(t_1-t_2))\right|^{-K}$ factors in their integrands respectively,
each of their characteristic behaviors can be captured by
$\int_{0}^\infty dt' \cos t' \left|\sinh(\pi k_BTt')\right|^{-K}$
and $\int_{0}^\infty dt' \sin t' \left|\sinh(\pi
k_BTt')\right|^{-K}$ with the latter dominating over the former
 for $(K\pi k_BT)\ll 1$.

 Comparison between $\widetilde{\eF}^{-}_1(\omega)$   and
$i\widetilde{\eH}^{-}_1(\omega)$ plotted in Fig.~\ref{fig:FH},  this observation and
 hence $\widetilde{\eA}(\omega)$ in the vicinity of
$\omega=\omega_0$ can be approximated as
\begin{equation}
\widetilde{\eA}(\omega)\approx
\frac{i}{2}\widetilde{\eH}^{-}_1(\omega_0-\omega)
\end{equation}
plus small corrections from other contributions including the tail
of the same function centered around $\omega=-\omega_0$:
$\widetilde{\eH}^{-}_1(\omega_0+\omega) $. Since
$\widetilde{\eH}^{\tilde{\epsilon}}_\zeta(\omega)$ is an {\em odd}
function of $\omega$ as we mentioned earlier, the dominant
contribution from $ \widetilde{\eH}^{-}_1(\omega_0-\omega)$  to
$\widetilde{\eA}(\omega)$  makes it nearly cross $\omega$ axis in
the vicinity of the Josephson frequency and thus suppressing the
contribution of  $\widetilde{\eA}(\omega_0)$ to the total noise.

\begin{figure}
\psfrag{F}{$\widetilde{\eF}^{-}_1(\omega)$}
\psfrag{H}{$\widetilde{i\eH}^{-}_1(\omega)$}
\psfrag{-2}{\small$-2$} \psfrag{-1}{\small$-1$}
\psfrag{1}{\small$1$} \psfrag{2}{\small$2$}
\psfrag{w}{$\omega/\omega_0$}
\includegraphics[width=0.4\textwidth]{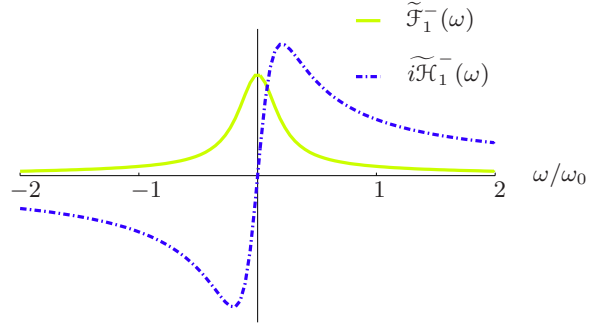}
\caption{Two dominant contributions to $\widetilde{\eA}$ as a
function of $\omega/\omega_0$ at temperature $T=0.5T_0$:
$\widetilde{\eF}^{-}_1(\omega)$ in (green) solid line
and $i\widetilde{\eH}^{-1}_1(\omega/\omega_0)$ in (blue) dotted
line. The width of the peaks are $2Kk_BT$.
$i\widetilde{\eH}^{-}_1(\omega)$ is already dominant
with stronger contribution at the tail even at this temperature.
As the temperature is further lowered, the dominance of
$i\widetilde{\eH}^{-}_1(\omega)$ over
$\widetilde{\eF}^{-}_1(\omega)$ becomes even more
prominent.} \label{fig:FH}
\end{figure}

We now turn to $\widetilde{\eB}$ which determines the magnitude of
the ``exchange''  term that bears the statistical angle
dependence. Given that only domain $R_2$, $\tilde{\epsilon}=+$
contributes to $\widetilde{\eB}$, the parts of the integrands that
strictly depend only on  $t_1$ and $t_2$ now becomes
$\cos\left(\omega_0(t_1+t_2)\right) \left|\sinh(\pi k_B
T(t_1-t_2))\right|^K$ and $\sin\left(\omega_0(t_1+t_2)\right)
\left|\sinh(\pi k_B T(t_1-t_2))\right|^K$ respectively for
$\widetilde{\eF}^{+}_2$ and $\widetilde{\eH}^{+}_2$. Since the
argument of the trigonometric function is the sum $(t_1+t_2)$
while the difference $(t_1-t_2)$ enters $\sinh$ function in a
non-singular fashion, both $\widetilde{\eF}^{+}_2(\omega)$ and
$\widetilde{\eH}^{+}_2(\omega)$ are of comparable magnitude away
from their centers $\omega=0$ (where
$\widetilde{\eF}^{+}_2(\omega=0)\neq 0$ (even function) while
$\widetilde{\eH}^{+}_2(\omega=0)=0$ (odd function)). Thus
$\widetilde{\eB}(\omega\sim\omega_0)$ can be appropriately
approximated as
\begin{equation}
\widetilde{\eB}(\omega) \approx
\widetilde{\eF}^{+}_2(\omega_0-\omega)
\end{equation}
which has a peak centered at $\omega=\omega_0$.

 \begin{figure}[ht]
 \subfigure[]
 {
\psfrag{S}{\footnotesize$\widetilde{S}(\omega/\omega_0)$}
\psfrag{S1by5}{\small$\widetilde{S}(\omega/\omega_0)$ for
$\nu\!=\!1/5$}
\psfrag{S2by5}{\small$\widetilde{S}(\omega/\omega_0)$ for
$\nu\!=\!2/5$} \psfrag{w}{\footnotesize$\omega/\omega_0$}
\psfrag{20}{\footnotesize$20$}\psfrag{10}{\footnotesize$10$}\psfrag{-10}{\footnotesize$-10$}
\psfrag{-20}{\footnotesize$-20$} \psfrag{0.5}{\footnotesize$0.5$}
\psfrag{1}{\footnotesize$1$}\psfrag{1.5}{\footnotesize$1.5$}
\includegraphics[width=0.37\textwidth]{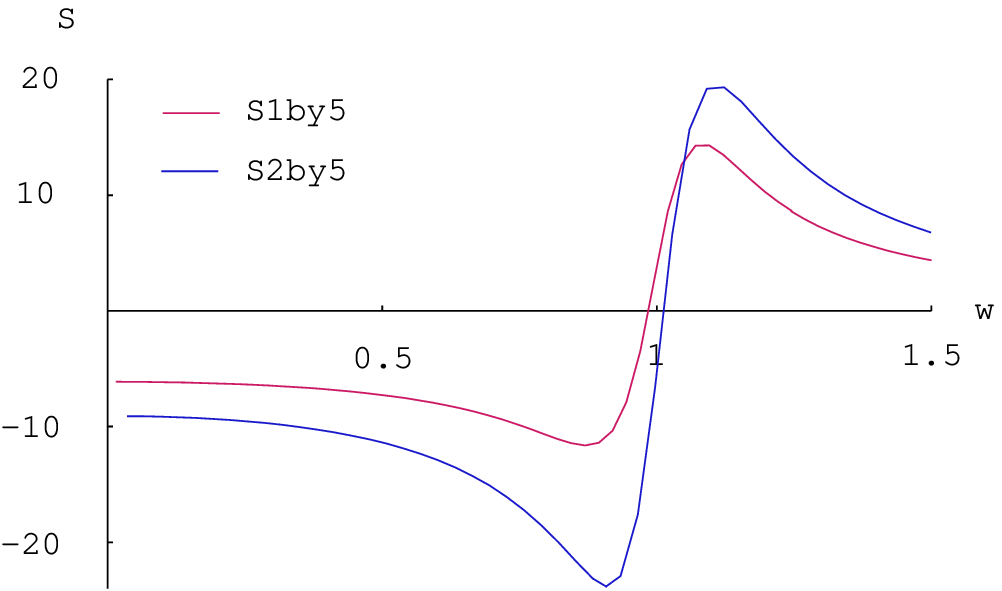}
} \subfigure[] { \psfrag{e}{$e$} \psfrag{e}{\footnotesize$\bar e$}
\psfrag{T}{\footnotesize$T/T_0$} \psfrag{1by5}{\footnotesize$\bar
e_{1/5}$} \psfrag{2by5}{\footnotesize$\bar e_{2/5}$}
\psfrag{0.5}{\footnotesize$0.5$} \psfrag{1}{\footnotesize$1$}
\psfrag{0.19}{\footnotesize$0.19$}
\psfrag{0.2}{\footnotesize$e^*(0.2)$}
\psfrag{0.21}{\footnotesize$0.21$}
\psfrag{0.22}{\footnotesize$0.22$}
\includegraphics[width=0.27\textwidth]{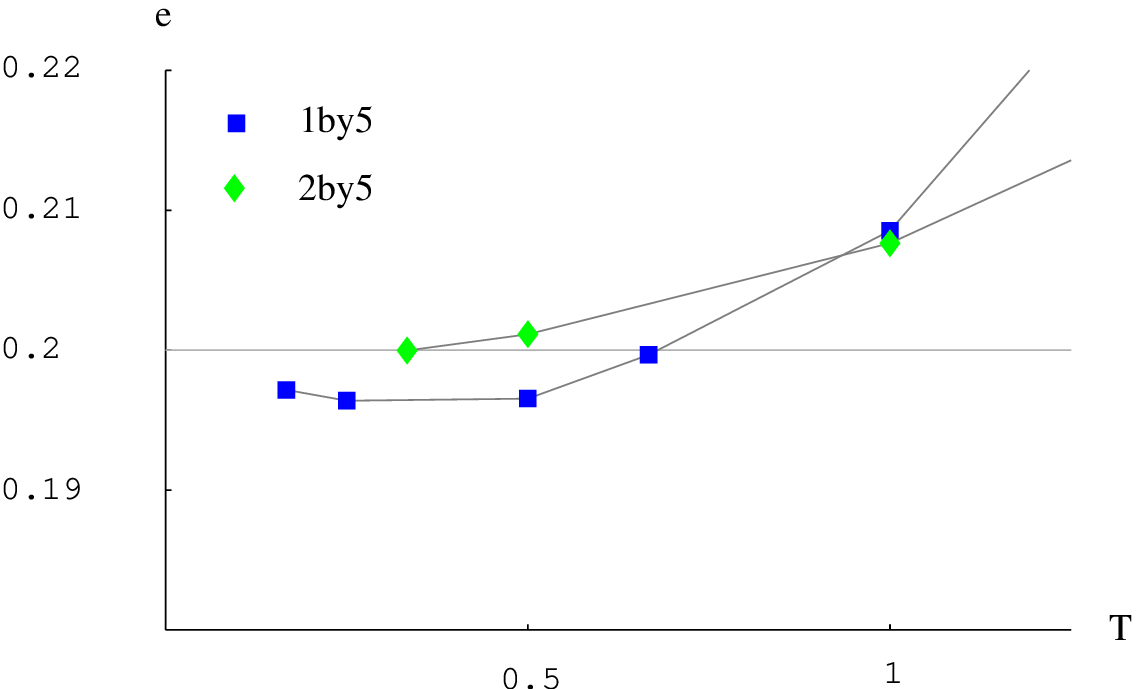}
} \caption{\label{fig:Sw-peak} (a)
$\widetilde{S}(\omega/\omega_0)$ for $\nu\!=\!1/5$ and
$\nu\!=\!2/5$ for $T=20\text{m}K,40 \text{m}K$ respectively with
$V\!\sim\!40\mu V$.
$\widetilde{S}(\omega/\omega_0)$ shows a peak in the vicinity of
the Josephson frequency $\omega_0$. At $\omega_0$,
$\widetilde{S}(\omega/\omega_0=1)$ is positive for the Laughlin
state $\nu=1/5$ and negative for the Jain state $\nu=2/5$. (b) The
effective charge $\bar e (T)$ in units of $e$ is determined by the
crossing points of (a) for $\nu\!=\!1/5$ and $\nu\!=\!2/5$. }
\end{figure}

From the preceding analysis, we can conclude that  the exchange
term plays a significant role in the measured net correlation
$\widetilde S(\omega/\omega_0\!=\!1)$ by shifting it to a positive
(negative) value for Laughlin (non-Laughlin) states since the
exchange term has $\cos\theta>0$ ($\cos\theta<0$) and the direct
term $\widetilde{\eA}(\omega)$ is strongly suppressed near
Josephson frequency. These sharp features not only provide a
direct measurement of the fractional charge associated with the
Josephson frequency but also allow a natural contrast between two
filling factors with the same charge but different statistics,
{\it e.g.} $\nu=1/5$ and $\nu=2/5$ as we show in
Fig.~\ref{fig:Sw-peak}(a). The information on statistics and
charge contained in the noise near Josephson frequency is
recapitulated in Fig.~\ref{fig:Sw-peak}(b) by introducing an
effective charge $\bar e \equiv\hslash\bar\omega/V$ from the
crossing point of $\widetilde S$  {\em i.e. } $\widetilde
S(\bar\omega/\omega_0)=0)$. The deviation between $\bar e$ for
$\nu\!=\!1/5$ and $\nu\!=\!2/5$ satisfying $\bar
e_{1/5}\!<\!e^*\!<\!\bar e_{2/5}$ over a broad temperature range
despite their common fractional charge $e^*\!=\!e/5$ is a
consequence of different statistical angles: $\theta\!=\!\pi/5$
(bunching) for $\nu=1/5$ and $\theta\!=\!3\pi/5$ (anti-bunching)
for $\nu\!=\!2/5$. Observation of such deviation will be a
definite signature for the existence of anyons.

We conclude this section by summing up the two main results which
show ways to extract the effect of fractional statistics:
$\widetilde S(\omega\!=\!0)=-|\widetilde
\eA_0|+\cos\theta|\widetilde \eB_0|$ for the zero frequency
Fourier spectrum plotted in Fig.~\ref{fig:Sw0}; $e^*$ defined from
$\widetilde{S}(\omega)$ for $\omega$ near Josephson frequency. The
first result can be used to measure the statistical angle in a
quantitative manner while the second result can be used to
determine the $\sgn(\cos\theta)$ and hence detect bunching versus
anti-bunching behavior.

 \section{Concluding remarks and multi-terminal noise experiments}
 \label{sec:conclusion}
 In summary, we have shown that the cross-current  noise in a
T-junction can be used to directly measure  the statistical angle
$\theta$ independently from the fractional charge, and
conceptually, from the filling factor $\nu$. The
normalized cross-current noise as a function of frequency $\omega$
assumes the form of a scaling function that depends on the
quantities $\omega/\omega_0$ and $\hbar\omega_0/kT$, where
'$\omega_0$' is the Josephson frequency, and all the statistical
dependence is contained in a simple $\cos\theta$ factor. The low
frequency results show that there is a quantitatively significant
statistics dependence of this noise. We have also shown how to
measure both fractional charge and statistics near the Josephson
frequency. Specifically, since the Josephson frequency explicitly
depends on the fractional charge, identifying its approximate
value based on where the finite frequency cross-current noise
drops to zero would provide a means of measuring fractional
charge. It must be noted that the fractional
charge, being a single-particle property, can be detected through
auto-current correlations, i.e., shot noise measurements. The zero frequency shot noise is proportional to the fractional charge in the weak tunneling limit~\cite{kane92l,picciotto97,saminadayar97,reznikov99}
while the finite frequency shot noise shows a peak at  the Josephson frequency\cite{chamon96}.

Plotting the actual frequency at which the cross-current noise
drops to zero as a function of temperature and comparing this
value to the expected Josephson frequency would give a measure of
the bunching versus anti-bunching behavior of the qp's arising
from statistics. A definitive means of measuring such statistical
behavior would be by comparing the cross-current noise of two FQH
states with the same fractional charge but different fractional
statistics, thus pinpointing the effect of topological order. We
remark that our results are independent of non-universal short
time physics.

 In this paper, we have considered a setup, the T-junction, consisting of quantum Hall bar with three terminals. In this setup, two quasiparticles (or a quasiparticle-quasihole pair) are created (or destroyed) at nearby places of one terminal (terminal 0 in Fig. 2). In the final state the two other terminals (1 and 2 in Fig. 2), which are widely separated, each detects just one quasiparticle. We have assumed throughout that the two quasiparticles are created at nearby points in terminal 0 separated by a distance $a$ small compared with the magnetic length. 
 Our calculation can be straightforwardly generalized for the situation in which $a$ is somewhat greater than the magnetic length.  However, as $a$ increases, various decoherence effects will hinder the detection of the statistical phase. In particular, when $a\gtrsim v\hslash/k_BT$ where $v$ is the speed of the propagating charge mode,  thermal fluctuation will overwhelm the coherent propagation of edge state between two points, and the interference effects associated with fractional statistics will become rapidly unobservable.

% In this paper, we have made the assumption of a single-point
%tunneling in that qp's are partitioned from a single point on the
%central edge state in the T-junction geometry into the two others.
%In reality, qp's may tunnel from two different points along the
%central edge determined by the closest locations to the two other
%edge states. The formalism developed in this paper can generalize
%our calculation to the situation where tunneling into the two edge
%states can occur from  two locations along the central edge state
%separated by a distance '$a$'. We expect cross-current
%correlations to remain robust as long as this spatial separation
%is less than a magnetic length. However, we expect the effect of
%statistics to decay once the separation length becomes greater
%than several magnetic lengths. 
%This behavior would be captured in
%the scaling function by an additional parameter $\omega_0a$.

Safi et.al\cite{safi01} have recently considered the identical
setup  for Laughlin states at $T=0$ and  obtained a series
expression for $\widetilde S(\omega\!=\!0)$ for the case $O$ alone
as a function of $\nu$ (note that for Laughlin states one cannot
distinguish $\nu$ from $K$ or $e^*/e$ or $\theta/\pi$). Indeed,
our finite temperature calculation shows that the total cross
current noise will be completely dominated by case $O$ as
$T\rightarrow 0$ which was the case of interest in
Ref.[\onlinecite{safi01}]. However, our study also shows that, at {\em
finite temperatures}, case $S$ brings in an explicit statistics
dependence in the form of $\cos\theta$ which can be distinguished
from the effect of the fractional charge and the scaling dimension
for non-Laughlin states.  Furthermore, even at relatively low
temperature at which most noise experiments involving QH systems
take place (see below), this statistics dependent contribution was
found to be of comparable magnitude as the statistics independent
contribution. It is rather encouraging to find  that a purely
quantum mechanical property such as statistics can in principle be
observed in our setup at non-zero temperatures. In fact it is
precisely  the finite temperature that assists processes that are
sensitive to statistics. A closely related setup is the four
terminal case studied (only for Laughlin states at $T=0$) by
Vishveshwara in Ref.[\onlinecite{vishveshwara03}] where the cross current
noise at equal times $t=0$ was calculated for the contribution
that involves all four terminals.  We should note that however,
even in this case, an actual measurement of the cross-correlation
will have contributions from correlation between three of the four
terminals which is precisely  $S(t)$ we have calculated
here. A number of interesting interferometers have been proposed to detect fractional statistics
\cite{jain93,chamon-freed-kivelson-sondhi-wen97,kane03,law06},  none of which has yet been realized experimentally, except possibly for the recent experiments of Ref.[\onlinecite{camino-prl,camino-T,zhou}]. Theoretical schemes to measure non-Abelian statistics have also been proposed.\cite{fradkin98,freedman05}

Here we have presented the calculation using the edge state theory constructed by Lopez
and Fradkin for Jain states~\cite{lopez99}. 
However, as long as one can limit oneself to a single propagating mode by either choosing the most relevant mode of hierarchical picture (see below) or by using Lopez-Fradkin picture which only has one propagating mode by design, the theory presented here is applicable and could indeed be used to compare predictions from different descriptions. 

 For the T-junction setup studied here, the hierarchical picture would imply multiple propagating edge modes for non-Laughlin states. 
%As we mentioned in the
%Introduction, alternative descriptions predict a somewhat
%different spectrum of qp's for non-Laughlin states.
In Ref.[\onlinecite{wen95}], Wen derived the effective theory for edge
states of  hierarchical FQH states following HaldaneÕs
hierarchical construction~\cite{haldane83}.
According to the hierarchical theory, the $\nu = 2/5$ FQH state is
generated by the condensation of quasiparticles on top of the $\nu
= 1/3$ FQH state. Thus the $2/5$ state contains two components of
incompressible fluids.  If we consider a special edge potential
such that the FQH state consists of two droplets, then in this
picture, one is the electron condensate with a filling fraction
$\nu_1=1/3$ and radius $r_1$, and the other is the quasiparticle
condensate (on top of the $1/3$ state) with a filling fraction
$\nu_2=1/15$ ( note $1/3 + 1/15 = 2/5$) and radius $r_2 < r_1$.
 Recently, one of us used this picture~\cite{kim06} to account for the superperiod Aharonov-Bohm oscillations observed in recent experiments~\cite{camino-prl,camino-T,zhou}  where the smooth edge potential could indeed have created two droplet situation.  
 On the other hand, if the edge is sharply defined so that there can be direct tunneling between descendant states ($2/5$ state for example), it is reasonable to consider the tunneling current being dominantly carried by the mode which has the smallest scaling dimension (that is ``most''  relevant in the renormalization group sense). In Wen's theory, although the primary edge qp mode carries $e^*\!=\!1/5$, $\theta/\pi\!=\!K\!=\!3/5$, it is actually the qp mode with $e^*\!=\!\theta/\pi\!=\!K=2/5$ that is most relevant~\cite{wen95}, which is quite different from the $e^*\!=\!1/5$, $\theta/\pi\!=\!3/5$, $K\!=\!1/10$ for the charge mode in Lopez-Fradkin picture.
In fact, given that our calculation is applicable to alternate pictures which come with different predictions, an experiment such as the one proposed here, where the charge and statistics of Jain state quasiparticles can seperately be identified, becomes all the more immediate.

Turning to experiments, some recent mesurements in a quantum
interferometer geometry were interpreted to imply fractional
statistics. An effective charge extracted in recent shot noise
experiments ~\cite{chung03} was somewhat larger than the minimum
quasiparticle charge, suggesting bunching behavior. Observation of
Aharonov-Bohm oscillations with a superperiod in an anti-dot
~\cite{camino-T,camino-prl,zhou} have also been ascribed to statistical effects.
However, the noise measurements we propose here will allow for a
direct and independent way of measuring both fractional statistics
and charge separately. Low frequency prediction have the advantage of yielding quantitatively significant statistics dependence of the noise.
Near the Josephson frequency, our prediction has the strength of
probing both charge and statistics simultaneously. The geometry
proposed and the required experimental conditions are all very
much within current capabilities. The T-junction setup is
reminiscent of the multi-lead geometry employed in effective
charge measurement, though, the issue of whether or not the
fabricated T-junction would form a single point contact would be
relevant. Typical parameters used in noise experiments
\cite{reznikov99,chung03} are bias voltages of $10 -150 \mu V$
(corresponding to a Josephson frequency $\omega_0 \sim 1$ GHz),
and temperatures of $10 - 100mK$. These parameters access the
ratios of $T/T_0$ shown here, making our proposal plausible.The
parameter regime in which the perturbative results for
cross-current correlations are valid depends on the inter-edge
tunneling amplitude and consequently on fabrication. The condition
(on applied voltage and temperature) for remaining in this regime
is that the ratio between tunneling current and maximum edge state
current $\nu e^2V/h$ be small, i.e., of order $0.1$.

A definitive measurement of anyonic statistics, such as the one
proposed here, would be one of the first instances of
experimentally establishing the existence of topological order
\cite{wen90,wen95}. Employing topological order has recently emerged as a possible 
route to decoherence
free quantum computing\cite{kitaev03}. Clear signatures of
fractional statistics would mark the very first steps towards
bringing some of these fascinating ideas to life.

{\bf Acknowledgments}:
This work was supported in part by the National
Science Foundation through the grants DMR 04-42537 (EK, ML, EF), DOE-MRL DEFG02-91-ER45439 (SV), by the Research Board of the University of Illinois (EF, ML), and by  a Stanford Institute for Theoretical Physics (EK). We  thank F.D.M. Haldane for illuminating discussions regarding exclusion statistics in the setup of this paper, and C. Chamon for many helpful comments (particularly on Klein factors). We also thank M. Heiblum and C. Marcus for discussions on noise experiments. We thank E. Ardonne for many helpful discussions.

\appendix

\section{Chiral and non-chiral boson commutation relations and correlation functions}
\label{ap:boson} In this appendix we quantize the chiral boson
theory and calculate its correlation function. We also extend this
calculation to the case of a neutral topological field that is
non-propagating.

The Lagrangian density  for a left/right moving free massless
chiral boson field $\phi_\pm(x\pm vt)$ propagating with speed $v$
is
    \begin{equation}
    {\mathcal  L}_\pm=\frac{g}{4\pi}\partial_x\phi_\pm(\pm\partial_t\phi_\pm-v\partial_x\phi_\pm)
    \label{eq:chiral-L}
    \end{equation}
with the corresponding action $S_0=\int d^2r {\mathcal L}$ where
$\vec{r}=(x,t)$ and $g$ is a positive real  parameter. Notice that
the Lagrangian is invariant under an arbitrary overall shift of
the field $\phi_\pm$, i.e.
    \begin{equation}
    \mathcal{L}_\pm[\phi_\pm(\vec{r})] = \mathcal{L}_\pm[\phi_\pm(\vec{r})+\alpha]
    \label{eq:shift}
    \end{equation}
with an arbitrary constant $\alpha$. Hence $\mathcal{L}_\pm$ is
independent of the zero mode of $\phi_\pm$.

Correct quantization of these chiral bosons can be achieved by the
following equal time commutation relations
    \begin{equation}
    [\phi_\pm(x,t), \pm\frac{g}{4\pi}\partial_x'\phi_\pm(x',t)]=\frac{1}{2}i\delta(x-x')
    \label{eq:chiral-comm}
    \end{equation}
where the factor of $\frac{1}{2}$ comes from the fact that
$\phi_\pm$ are non-local fields with respect to the non-chiral
boson $\phi\equiv\phi_-+\phi_+$ occupying only half the phase
space of $\phi$.  The non-chiral boson $\phi$ is a true local
field and it can be quantized in the usual way. We will find below
that Eq.~\eqref{eq:chiral-comm} is indeed the correct commutation
relations consistent with the commutation relations of non-chiral
boson.
To find the non-chiral boson Lagrangian using a path integral, we
need another (dual) non-chiral bosonic field $\theta$ to change
the basis from $\phi_-$ and $\phi_+$ to $\phi$ and $\theta$. Hence
we define
    \begin{equation}
    \theta\equiv \phi_- - \phi_+
    \label{eq:dual}
    \end{equation}
and write $\phi_\pm$ in terms of $\phi$ and $\theta$ as the
following.
    \begin{equation}
    \phi_+=\frac{1}{2}(\phi-\theta),\qquad
    \phi_-=\frac{1}{2}(\phi+\theta)
    \label{eq:phi-pm}
    \end{equation}
Then the total Lagrangian for two fields can be written in terms
of $\phi$  and $\theta$ as the following
    \begin{equation}
    \mathcal{L}_+ +\mathcal{L}_-
    =-\frac{g}{8\pi}[\partial_x\phi\partial_t\theta+\partial_x\theta\partial_t\phi+v(\partial_x\phi)^2+v(\partial_x\theta)^2]
    \end{equation}
and the path-integral for two chiral fields can be written as a
path-integral for two non-chiral fields to give
    \begin{equation}
    \begin{split}
    \mathcal{Z}
    &=\int{\EuScript D}\phi_-\int{\EuScript D}\phi_+ e^{i\int d^2r [\mathcal{L}_++\mathcal{L}_-]}\\
    &=\int{\EuScript D}\phi\int{\EuScript D}\theta e^{-i\frac{g}{8\pi}\int d^2r [v(\partial_x\theta)^2+2\partial_t\phi\partial_x\theta+v(\partial_x\phi)^2]}.
    \end{split}
    \end{equation}
Now we can integrate $\theta$ out from the above equation
    \begin{equation}
    \mathcal{Z}=\int{\EuScript D}\phi e^{\D i\int d^2r\frac{g}{8\pi}[\frac{1}{v}(\partial_t\phi)^2-v(\partial_x\phi)^2]}
    \end{equation}
to obtain the Lagrangian for the non-chiral field $\phi$
    \begin{equation}
    \mathcal{L}[\phi] =\frac{g}{8\pi} [\frac{1}{v}(\partial_t\phi)^2-v(\partial_x\phi)^2].
    \end{equation}
We can find the canonical conjugate of $\phi$, $\Pi_\phi$ in the
usual way  to be
    \begin{equation}
    \Pi_\phi\equiv\frac{\delta\mathcal{L}}{\delta\partial_t\phi}=\frac{1}{v}\frac{g}{4\pi}\partial_t\phi
    \end{equation}
with the equal time commutation relation
    \begin{equation}
    [\phi(x),\Pi_\phi(x')]=i\delta(x-x').
    \label{eq:phi-Pi-comm}
    \end{equation}
However, since $\phi=\phi_-(x-vt)+\phi_+(x+vt)$
    \begin{equation}
    \begin{split}
    \partial_t\phi
    &=\partial_t(\phi_-+\phi_+)\\
    &=-v\partial_x(\phi_-+\phi_+)\\
    &=-v\partial_x\theta
    \end{split}
    \end{equation}
which means $\partial_x\theta$ is proportional to the canonical
conjugate $\Pi_\phi$:
    \begin{equation}
    \Pi_\phi=-\frac{g}{4\pi}\partial_x\theta
    \end{equation}
and hence Eq.~\eqref{eq:phi-Pi-comm} can be rewritten as
    \begin{equation}
    [\phi(x),\partial_{x'}\theta(x')]=-i\frac{4\pi}{g}\delta(x-x').
    \label{eq:phi-theta-comm}
    \end{equation}
From Eq.~\eqref{eq:phi-pm} and Eq.~\eqref{eq:phi-theta-comm} one
can check that Eq.~\eqref{eq:chiral-comm} indeed is the correct
way of quantizing the chiral modes.

Now integrating both sides of Eq.~\eqref{eq:chiral-comm} with
respect to $x'-x$ and noting that
$\partial_{x'-x}[\phi_\pm(x),\phi_\pm(x')]~\!=\!~[\phi_\pm(x),\partial_{x'}\phi_\pm(x')]~\!-\!~[\partial_x\phi_\pm(x),\phi_\pm(x')]$,
one finds the following {\em equal time} commutation relations for
{\em left/right} moving bosons $\phi_\pm$
    \begin{equation}
    \begin{split}
    [\phi_+(x),\phi_+(x')]&=-i\frac{\pi}{g}\sgn(x-x')\\
    [\phi_-(x),\phi_-(x')]&=i\frac{\pi}{g}\sgn(x-x').
    \end{split}
    \label{eq:chiral-comm-sgn}
    \end{equation}

To calculate the chiral boson propagator, we first define a
generating functional:
    \begin{equation}
    \begin{split}
    Z[a]&=\int\!\!\mathcal{D}\phi_\pm\ e^{i\int d^2 r[\mathcal{L}_\pm[\phi_\pm]+a\phi_\pm}]\\
        &=Z[0] \ e^{-\frac{1}{2}\int d^2r\int d^2r' a(r)M_\pm^{-1}(r-r')a(r')}
    \end{split}
    \end{equation}
where we defined the differential operator associated with the Lagrangian $\mathcal{L}_\pm$
    \begin{equation}
    M_\pm\equiv i\frac{g}{2\pi}\partial_x(\pm\partial_t-v\partial_x).
    \end{equation}
Now the propagator is nothing but the inverse of $M$ since
    \begin{equation}
    \begin{split}
    G_\pm(\vec{r}-\vec{r}')&=\langle\phi_\pm(\vec{r})\phi_\pm(\vec{r}')\rangle\\
    &= -\left. \frac{\delta \ln Z[a]}{\delta a(\vec{r})\delta a(\vec{r}')}\right|_{ a=0}\\
    &= M_\pm^{-1}(\vec{r}-\vec{r}'),
    \end{split}
    \end{equation}
and can be calculated using a Fourier transformation with the
proper epsilon prescription for  a time ordered propagator
(substituting $\omega$ with $\omega+i\epsilon\sgn(\omega)$ in the
limit $\epsilon\rightarrow 0^+$):
    \begin{equation}
    \begin{split}
    G_\pm(\vec{r})&=\int\frac{dkd\omega}{(2\pi)^2}\frac{2\pi}{ig}\frac{e^{i(kx-\omega t)}}{ik[\mp i(\omega+i\epsilon\sgn(\omega))-ivk]}\\
    &=i\int\frac{dkd\omega}{2\pi}\frac{1}{g}\frac{e^{-i(\omega t\pm kx)}}{k(\omega+ i\epsilon\sgn(\omega)-vk)}\\
    &=\frac{1}{g}\int\frac{dk}{k}[e^{-ik(vt\pm x)-k\tau_0}\Theta(k)\Theta(t)\\
    &\qquad\qquad\quad-e^{ik(vt\pm x)+k\tau_0}\Theta(-k)\Theta(-t)].
    \end{split}
    \label{eq:Gb}
    \end{equation}
Here we introduced a UV (short distance) cutoff $\tau_0$ and used
the fact that the $\omega$ contour integral has to be closed in
the lower(upper) half plane for $t>0$ ($t<0$) and $ \sgn(\omega)=
\sgn(k)$ near the pole at $\omega=vk$. Notice that the left
(right) moving propagator $G_+$ ($G_-$) indeed depends only on
$vt+x$ ($vt-x$), i.e. $\phi_\pm(x,t)=\phi_\pm(vt\pm x)$. It is
clear from Eq.~\eqref{eq:Gb} that $G(r)$ has a logarithmic
divergence in the small $k$ (IR) limit. However, one can define a
finite quantity by subtracting the equal position propagator
$G(0)$, which also has a logarithmic divergence, and obtain the
following
    \begin{widetext}
    \begin{equation}
    \begin{split}
    \tilde{G}_\pm(\vec{r})&\equiv G_\pm(\vec{r})-G_\pm(0)\\
    &=\frac{1}{g}\int\frac{dk}{k}[\{e^{-ik(vt\pm x)-k\tau_0}-1\}\Theta(k)\Theta(t)-\{e^{ik(vt\pm x)+k\tau_0}-1\}\Theta(-k)\Theta(-t)]\\
    &=-\frac{1}{g}\ln\left[\frac{\tau_0+i\sgn(t)(vt\pm x)}{\tau_0}\right].
    \end{split}
    \label{eq:chiralG}
    \end{equation}
    \end{widetext}

At finite temperatures $T$,  the regulated and time ordered propagator  can be calculated from Eq.~\eqref{eq:chiralG} via conformal mapping     to take the following form 
    \begin{equation}
    \tilde{G}_\pm(x',t) = -\frac{1}{g}\ln\left[\frac{\frac{\pi}{\beta}\tau_0}{\sin\left(\frac{\pi\tau_0}{\beta}+i\sgn(t)\frac{\pi}{\beta}(vt\pm x)\right)} \right],
    \label{eq:finiteT-G}
    \end{equation}
where $\beta=1/(k_B T)$ is the inverse temperature with $k_B$
being the Boltzmann constant.

In the non-propagating limit of $v_N\rightarrow0^+$,
Eq.~\eqref{eq:chiral-L} becomes the Lagrangian for a topological
field\cite{lee-wen98,lopez99} as the following
    \begin{equation}
    {\mathcal  L}_N=\frac{g_N}{4\pi}\partial_x\phi_N\partial_t\phi_N
    \label{eq:neutralL}
    \end{equation}
where $g_N$ can be either positive or negative real number.
$\mathcal{L}_N$ is still invariant under an arbitrary overall
shift of the topological field $\phi_N$ and the propagator for the
topological field can be calculated in the similar way as for the
chiral field:
    \begin{equation}
    \begin{split}
    &\langle T\phi_N(\vec{r})\phi_N(0)\rangle\\
    &=-\frac{i}{g_N}\int\frac{dkd\omega}{2\pi}\frac{e^{i(kx-\omega t)}}{k\omega}\\
    &=-\frac{i}{g_N}\frac{1}{2\pi}(-i\pi\Theta(t)+i\pi\Theta(-t))(i\pi\Theta(x)-i\pi\Theta(-x))\\
    &=-\frac{i}{g_N}\frac{\pi}{2}\sgn(t)\sgn(x)\\
    &=\lim_{v_N\rightarrow 0^+}\frac{i}{g_N}\frac{\pi}{2}\sgn(t)\sgn(v_Nt+x),
    \end{split}
    \label{eq:neutralG}
    \end{equation}
where the limit should be taken at the end of any calculation that
uses $\langle\phi_N(\vec{r})\phi_N(0)\rangle$.

\section{Contour ordered (Keldysh) propagators for chiral bosons}
\label{ap:keldysh}

\begin{figure}[b]
    \psfrag{t}{forward $\eta=+$}
    \psfrag{b}{backward $\eta -$}
    \includegraphics[width=0.3\textwidth]{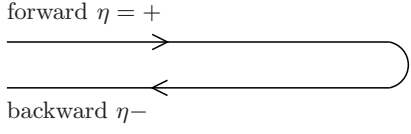}
    \caption{The Keldysh contour from $t=-\infty$  to $t=\infty$ along forward branch ($\eta=+$), and back to $t=-\infty$ along backward branch ($\eta=-$). }
    \label{fig:contour}
    \end{figure}
Here we derive the contour ordered Green functions for chiral
bosons  described by the Lagrangian Eq.~\eqref{eq:chiral-L} and
topological modes described by Eq.~\eqref{eq:neutralL}.
Time ordered propagator can be thought of as a sum of two terms
depending on the sign of $(t-t')$:
\begin{equation}
\begin{split}
&\langle T \phi(x,t)\phi(x',t')\rangle\\
&=\Theta(t\!-\!t')\langle\hat{\phi}(x,t)\hat{\phi}(x',t')\rangle
+\Theta(t'\!-\!t)\langle\hat{\phi}(x',t')\hat{\phi}(x,t)\rangle.
\label{eq:Tordering}
\end{split}
\end{equation}
In the same way, a contour ordered propagator is determined by the
order in which different times appear on the contour sketched in
Fig.~\ref{fig:contour}
    \begin{widetext}
    \begin{equation}
    G_{\eta,\eta'}(x-x', t-t')=\langle T_K \phi(x,t^\eta)\phi(x',t'^{\eta'})\rangle\equiv
    \left\{\begin{split}
        \langle \phi(x,t^\eta)\phi(x',t'^{\eta'})\rangle& \quad \text{for}\ t^\eta >_K t'^{\eta'}\\
        \langle \phi(x',t'^{\eta'})\phi(x,t^\eta)\rangle& \quad \text{for}\ t'^{\eta'}>_K t^\eta
        \end{split} \right.
    \end{equation}
    \end{widetext}
where we used the expression $t>_Kt'$ to denote $t$ appearing
later than $t'$ in the contour. Clearly  when both times occur in
the forward branch ($\eta=\eta'=+$), contour ordering is the same
as time ordering and when both times occur in the backward branch
($\eta=\eta'=-$), contour ordering is  the same as anti-time
ordering. On the other hand, when $t$ is on the backward branch
while $t'$ is on the forward branch ($\eta=-$, $\eta'=+$),
obviously $t$ is occurring later in the contour and the exact
opposite is the case for $\eta=+$, $\eta'=-$.
Comparing the time ordered propagator Eq.~\eqref{eq:chiralG} with
Eq.~\eqref{eq:Tordering} and setting $v=1$
    \begin{equation}
    \langle\hat{\phi}_{\pm}(x, t)\hat{\phi}_\pm(0, 0)\rangle = -\frac{1}{g}\ln\left[\frac{\tau_0+i(t\pm x)}{\tau_0}\right]
    \end{equation}
and combining this with the above observation, we arrive at the
following contour ordered propagators for the right moving chiral
boson $\phi_-$:
    \begin{widetext}
    \begin{equation}
    \begin{split}
    G_{-+}(x, t)&=\langle\phi_-(x, t)\phi_-(0, 0)\rangle= \frac{1}{g}\ln\left[\frac{\tau_0+i(t-x)}{\tau_0}\right]\\
    G_{+-}(x, t)&=\langle\phi_-(0, 0)\phi_-(x, t)\rangle= \frac{1}{g}\ln\left[\frac{\tau_0-i(t-x)}{\tau_0}\right]\\
    G_{++}(x, t)&=\langle T \phi_-(x, t)\phi_-(0, 0)\rangle= \frac{1}{g}\ln\left[\frac{\tau_0+i\sgn(t-0)(t-x)}{\tau_0}\right]\\
    G_{--}(x, t)&=\langle T \phi_-(0,0)\phi_-(x, t)\rangle= \frac{1}{g}\ln\left[\frac{\tau_0-i\sgn(t)(t-x)}{\tau_0}\right],
    \end{split}
    \end{equation}
    \end{widetext}
which can be written in the following compact form
    \begin{equation}
    G_{\eta,\eta'}(x, t) = -\frac{1}{g}\ln\left[ \frac{\tau_0+i\chi_{\eta,\eta'}(t)(t-x)}{\tau_0}\right]
    \label{eq:K-GzeroT}
    \end{equation}
where we introduced
    \begin{equation}
    \chi_{\eta, \eta'}(t)\equiv \frac{\eta+\eta'}{2}\sgn(t)-\frac{\eta-\eta'}{2}.
    \label{eq:chi}
    \end{equation}
Going through the similar analysis for the the neutral mode we
derive the following contour ordered propagator
    \begin{equation}
    G_{\eta, \eta'}^N(x,t) =\lim_{v_N\rightarrow 0} -i\frac{\pi}{2}\chi_{\eta, \eta'}(t)\sgn(v_Nt - x)
    \label{eq:K-neutralG}
    \end{equation}
Notice that both $G_{-+}(x, 0)-G_{+-}(x, 0)$ and $G_{-+}^N(x,
0)-G_{+-}^N(x, 0)$ properly yield appropriate commutation
relations for $\phi_-$ and $\phi_N$ as one would expect.
At finite temperatures, Eq.~\eqref{eq:K-GzeroT} becomes
    \begin{widetext}
    \begin{equation}
    G_{\eta,\eta'}(x, t) = -\frac{1}{g}\ln\left[ \frac{\D\sin\left(\frac{\pi\tau_0}{\beta}+i\chi_{\eta,\eta'}(t)\frac{\pi}{\beta}(t\!-\!x)\right)}{\D\frac{\pi\tau_0}{\beta}}\right].
    \label{eq:K-GfiniteT}
    \end{equation}
    \end{widetext}
However,  since the neutral mode is a non-propagating mode its
propagator is independent of the temperature.

Finally, in the limit of $x\rightarrow 0^-$ and
$\tau\rightarrow 0$, Eq.~\eqref{eq:K-GfiniteT} and
Eq.~\eqref{eq:K-neutralG} becomes,

    \begin{align}
    G_{\eta, \eta'}(0^-, t) &= -\frac{1}{g} \ln\left|\frac{\sinh\!\frac{\pi t}{\beta}}{\frac{\pi\tau_0}{\beta}}\right| - i\frac{1}{g}\frac{\pi}{2}\chi_{\eta, \eta'}(t)\sgn(t)\nonumber\\
    G_{\eta, \eta'}^N(0^-,t) & =\lim_{v_N\rightarrow 0^+} \left[-i\frac{1}{g}\frac{\pi}{2}\chi_{\eta, \eta'}(t)\sgn(v_Nt)\right] \\
        &=  -i\frac{1}{g}\frac{\pi}{2}\chi_{\eta, \eta'}(t)\sgn(t)\nonumber
    \end{align}
where we used
    \begin{equation}
    \ln(it)=  \ln (i\sgn(t))+\ln|t| =\ln|t| +i\frac{\pi}{2}\sgn(t)
    \end{equation}
and  $\lim_{v_N\rightarrow 0^+} \sgn(v_Nt)  = \sgn(t)$. Also it is
quite clear from the derivation that $G_{\eta, \eta'} (x, t
) = G_{\eta', \eta}(-x, -t)$.

\section{Multiple vertex function correlator}
\label{ap:vertex} 
Consider the vertex operators $V(q,\vec{r}) =
e^{iq\phi(\vec{r})}$. Here we calculate  the correlation function
of $N$ vertex operators $\langle V({q_1},\vec{r}_1)\cdots
V({q_N},\vec{r}_N)\rangle$ with respect to the free boson
Lagrangian Eq.~\eqref{eq:chiral-L}
    \begin{equation}
    \langle V({q_1},\vec{r}_1)\cdots V({q_N},\vec{r}_N)\rangle
    \equiv\frac{\displaystyle \int\!\!\mathcal{D}\phi\  e^{iS_0[\phi]}\  e^{i\sum_{j=1}^N   q_j\phi(\vec{r}_j)}}{\displaystyle\int\!\!\mathcal{D}\phi \ e^{-S_0[\phi]} }.
    \end{equation}
Clearly this correlator will have the symmetry
Eq.~\eqref{eq:shift} only when
\begin{equation}
\sum_{j=1}^N q_j=0 \label{eq:neutrality}
\end{equation}
which is a charge neutrality condition whose important consequence
will be shown below. Introducing a source term
    \begin{equation}
     a(\vec{r}) = \sum_{j=1}^Nq_j\delta^{(2)}(\vec{r}-\vec{r}_j),
    \label{eq:source}
    \end{equation}
one can calculate the correlator using path integral as follows:
    \begin{equation}
    \begin{split}
    \langle V({q_1},\vec{r}_1)&\cdots V({q_N},\vec{r}_N)\rangle\\
    &= \frac{\displaystyle \int\!\!\mathcal{D}\phi\  e^{iS_0[\phi]}\  e^{i\int d^2r a(r)\phi(r)}}{\displaystyle\int\!\!\mathcal{D}\phi \ e^{iS_0[\phi]} }\\
    &=e^{-\frac{1}{2}\int\! d^2r\int\! d^2r'  a(\vec{r})G(\vec{r}-\vec{r}') a(\vec{r}')}\\
    &=e^{-\frac{1}{2}\sum_{j,k=1}^Nq_jq_k G(\vec{r}_j-\vec{r}_k)}.
    \end{split}
    \label{eq:vertex-corr}
    \end{equation}
However the exponent can be separated into two parts since
    \begin{equation}
    \begin{split}
    \sum_{j,k=1}^N &q_j q_k  G(\vec{r}_j-\vec{r}_k)\\
    &=\sum_{j,k=1}^N q_j q_k[G(\vec{r}_j-\vec{r}_k)-G(0)+G(0)]\\
    &=(\sum_{j=1}^N q_j)^2 G(0) + \sum_{j,k=1}^N q_j q_k [G(\vec{r}_j-\vec{r}_k)-G(0)]\\
    &=(\sum_{j=1}^N q_j)^2 G(0) + 2\sum_{j>k=1}^N q_j q_k [G(\vec{r}_j-\vec{r}_k)-G(0)].
    \end{split}
    \end{equation}
Notice though $G(0)$ has IR divergence, as it was pointed out in
Appendix~\ref{ap:boson}, combined with the charge neutrality
condition Eq.~\eqref{eq:neutrality} the correlation function
becomes
\begin{equation}
\langle V({q_1},\vec{r}_1)\cdots
V({q_N},\vec{r}_N)\rangle=e^{-\!\sum_{k>j=1}^N q_jq_k
\tilde{G}(\vec{r}_j-\vec{r}_k)} \label{eq:neutral-vertex-corr}
\end{equation}
where the finite subtracted boson propagator
$\tilde{G}(\vec{r})\equiv G(\vec{r})-G(0)$ naturally shows up.

\section{Unitary Klein factors}
\label{ap:klein}
Here we derive the unitary Klein factor algebra\cite{guyon02} used
in Section~\ref{sec:setup} and their contour ordering. Anyonic
exchange statistics between qp's of different edges $l$ and $m$
defined in Eq.~\eqref{eq:l-vertex} requires
    \begin{equation}
    \psi_l^\dagger\psi_m^\dagger=e^{\pm i \theta}\psi_m^\dagger\psi_l^\dagger.
    \label{eq:l-vertex-exchange}
    \end{equation}
However since boson fields commute between different edges, the
statistical phase should come solely from the Klein factor
algebra. Hence if we define real parameters $\alpha_{lm}$ by
    \begin{equation}
    F_lF_m=e^{-i\alpha_{lm}}F_mF_l,
    \label{eq:Flm-exchange}
    \end{equation}
the magnitude of $\alpha_{lm}$ follows immediately from comparing
Eq.~\eqref{eq:l-vertex-exchange} with the following equation
    \begin{equation}
    \psi_l^\dagger\psi_m^\dagger=e^{-i\alpha_{lm}}\psi_m^\dagger\psi_l^\dagger
    \end{equation}
to be
    \begin{equation}
    \alpha_{lm}=\pm\theta=-\alpha_{ml}
    \label{eq:alpha-lm}
    \end{equation}
leaving the sign of $\alpha_{lm}$ still arbitrary.
\begin{figure}[h]
\psfrag{0}{\small$0$} \psfrag{1}{\small$1$} \psfrag{2}{\small$2$}
\psfrag{x=0}{\tiny$x\!=\!0$} \psfrag{x0=0}{\tiny$x_0\!=\!0$}
\psfrag{a}{(a)} \psfrag{b}{(b)}
\includegraphics[width=0.47\textwidth]{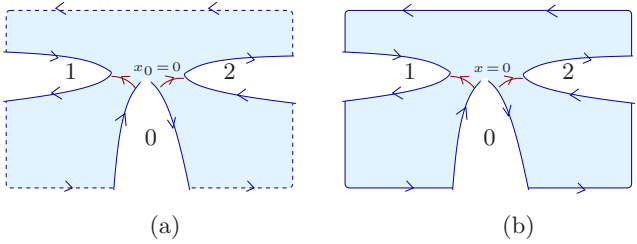}
\caption{Two configurations with the same topology: (a)Three edges
are disconnected and each are described by $\varphi_l$ with
$l=0,1,2$.  The tunneling to edges $l=1,2$ occurs at
$(x_0\!=\!-a/2\!\equiv\!X_1,x_1\!=\!0)$ and $
(x_0\!=\!a/2\!\equiv\! X_2, x_2\!=\!0)$ respectively. (b) All
three edges are connected with an open contour starting at $x=0$
which is described by a single chiral boson $\varphi$. For the
tunneling processes to edges $l=1,2$, qp's hop out from
$x\!=\!-a/2\!\equiv\!X_1$ and  $x\!=\!a/2\!\equiv\! X_2$ respectively. We note in passing that since the choice of the origin is arbitrary and do not affect the end result, we made a convenient choice.} \label{fig:klein}
\end{figure}
To consistently fix the correct sign, one must compare the system
which is equivalent to Fig.~\ref{fig:klein} (a) with the same
topology but where edges are connected with an open contour that
brings all the edges into one as in Fig.~\ref{fig:klein}(b)
following \textcite{guyon02}. Klein factors are not necessary in
the latter system because the commutation relations of tunneling
operators are enforced by the single chiral bosonic field $\phi$.

One way of determining Klein factors for case (a) is to compare
the relative phases $e^{-i\alpha_{lm}}$ of Fig.~\ref{fig:klein}
(a) with the phases $e^{i\theta\sgn(X_l-X_m)}$ of Fig.~\ref{fig:klein} (b)
where $X_l$ and $X_m$ are now abscissas of the same edge measured from the point $x=0$ (see the caption of Fig.~\ref{fig:klein}
).
Another way is to look at the commutation relations of different
tunneling operators $\hat{V}_l$ and to require  that the
commutator of the  tunneling operators of the disconnected system
give the same results as the connected one.

These two approaches give the same results for the Klein factors
and we will choose to detail the second approach here. For the
connected system of Fig.~\ref{fig:klein} (b), there is no need for
the Klein factors since the tunneling operators at different
positions will commute due to chirality (as long as the tunneling
paths do not cross). Hence we will set the constraint for
$\alpha$'s by requiring $\hat{V}_1$ and $\hat{V}_2$  in the case
of interest Fig.~\ref{fig:klein} (a) to commute. First, by
comparing the exchange of $\psi(x)\psi(x')$ with that of
$\psi^\dagger(x)\psi(x')$ using the CBH formula, one finds that
the definition of Eq.~\eqref{eq:Flm-exchange} implies
    \begin{equation}
    F_lF_m^{-1}=e^{i\alpha_{lm}}F_m^{-1}F_l.
    \label{eq:Flm-conj-exchange}
    \end{equation}
With Eqs.~\eqref{eq:Flm-exchange} and \eqref{eq:Flm-conj-exchange}
we can exchange $\hat{V}_1$ and $\hat{V}_2$ at equal time as the
following
    \begin{widetext}
    \begin{equation}
    \begin{split}
    \hat{V}_1\hat{V}_2
    &=F_0F_1^{-1}F_0F_2^{-1} e^{i\varphi_0(x_0=X_1)}e^{i\varphi_0(x_0=X_2)}e^{-i\varphi_1(x_1=0)}e^{-i\varphi_2(x_2=0)}\\
    &=F_0F_2^{-1}F_0F_2^{-1}e^{i(\alpha_{02}+\alpha_{10}+\alpha_{21})}e^{i\theta\sgn(X_1-X_2)} e^{i\varphi_0(X_2)}e^{i\varphi_0(X_1)} e^{-i\varphi_2(0)}e^{-i\varphi_1(0)}\\
    &=e^{i(\alpha_{02}+\alpha_{21}+\alpha_{10})}e^{i\theta\sgn(X_1-X_2)}\hat{V}_2\hat{V}_1
    \end{split}
    \label{eq:Vjk-exchange}
    \end{equation}
    \end{widetext}
where we used the fact that $[\varphi_l,\varphi_m]=0$  for $l\neq
m$ and
\begin{equation}
 e^{i\varphi_0(x_0=X_1)}e^{i\varphi_0(x_0=X_2)}
 =e^{i\theta\sgn(X_1-X_2)} e^{i\varphi_0(X_2)}e^{i\varphi_0(X_1)}.
\end{equation}
The second equality followed from qp exchange within edge $0$
given by Eq.~\eqref{eq:angle} and series of Klein factor exchange
as the following:
    \begin{equation}
    \begin{split}
    F_0F_1^{-1}F_0F_2^{-1}
    &=F_0F_1^{-1}F_2^{-1}F_0 e^{i\alpha_{02}}\\
    &=F_0F_2^{-1}F_1^{-1}F_0e^{i\alpha_{02}-i\alpha_{12}}\\
    &=F_0F_2^{-1}F_0F_1^{-1}e^{i\alpha_{02}-i\alpha_{12}+i\alpha_{10}}\\
    &=F_0F_2^{-1}F_0F_1^{-1}e^{i(\alpha_{02}+\alpha_{21}+\alpha_{10})}.
    \end{split}
    \end{equation}
From Eq.~\eqref{eq:Vjk-exchange}, requiring $\hat{V}_1$ and
$\hat{V}_2$ to commute as they would in a single connected edge,
results in the constraint:
\begin{equation}
\alpha_{02}+\alpha_{21}+\alpha_{10}+\theta\sgn(X_1-X_2)= 0,
\end{equation}
 which allows us to define $\alpha_{02}=\alpha_{21}=-\alpha_{10}=\theta$ since  $X_2\!-\!X_1\!=\!a\!>\!0$(See the caption of Fig.~\ref{fig:klein}).
Further we can conclude
\begin{equation}
F_0F_1^{-1}F_0F_2^{-1}=F_0F_2^{-1}F_0F_1^{-1}e^{i\theta}.
    \label{eq:F0Fj-exchage}
\end{equation}

Now we determine the contour ordered Klein factor  ``propagator''
$\langle T_K
(F_0F_1^{-1})^\epsilon(t^\eta)(F_0F_2^{-1})^{\epsilon'}(t'^{\eta'})\rangle_0
$ based on Eq.~\eqref{eq:F0Fj-exchage}.  We first note contour
ordering means that $(F_0F_1^{-1})^\epsilon(t^\eta)$ and
$(F_0F_2^{-1})^{\epsilon'}(t'^{\eta'})$ need to be exchanged
whenever $t'^{\eta'}\!\!>_Kt^{\eta}$, i.e.
    \begin{widetext}
    \begin{equation}
    \langle T_K  (F_0F_1^{-1})^\epsilon(t^\eta)(F_0F_2^{-1})^{\epsilon'}(t'^{\eta'})\rangle_0 =\left\{
    \begin{split}
    &\langle (F_0F_1^{-1})^\epsilon(t^\eta)(F_0F_2^{-1})^{\epsilon'}(t'^{\eta'})\rangle \quad \text {for}\  t^\eta>_Kt'^{\eta'}\\
    &\langle (F_0F_2^{-1})^{\epsilon'}(t'^{\eta'}) (F_0F_1^{-1})^\epsilon(t^\eta)  \rangle  \quad \text {for}\  t'^{\eta'}\!\!>_Kt^{\eta}
    \end{split}
    \right.
    \label{eq:def-F-klein}
    \end{equation}
    \end{widetext}
From Eq.~\eqref{eq:F0Fj-exchage}, it is clear that a consistent
way of defining the contour ordered propgator is to assign a
relative phase factor of $e^{-i\theta\sgn(X_1-X_2)}$ between
$t'^{\eta'}\!\!>_Kt^{\eta}$ case (in which the Klein factors need
to be exchanged for contour ordering) and $t^\eta>_Kt'^{\eta'}$
case(in which no exchange is neccesary). Further in order for
$\langle T_K \hat{V}_1(t^\eta)\hat{V}_2(t'^{\eta'})\rangle =
\langle T_K\hat{V}_2(t'^{\eta'}) \hat{V}_1(t^\eta)\rangle$ to
hold, which would be the case for the connected edge and which
connects us back to the basis for the derivation of
Eq.~\eqref{eq:F0Fj-exchage}, we require
    \begin{widetext}
    \begin{equation}
    \langle T_K(F_0F_1^{-1})^{\epsilon}(t^\eta) (F_0F_2^{-1})^{\epsilon'}(t'^{\eta'}) \rangle = \langle T_K(F_0F_2^{-1})^{\epsilon'}(t'^{\eta'}) (F_0F_1^{-1})^{\epsilon}(t^\eta) \rangle .
    \end{equation}
    \end{widetext}
This last requirement can be satisfied by assigning half of the
relative phase factor   $e^{-i\theta\sgn(X_1-X_2)}$ to each case
of Eq.~\eqref{eq:def-F-klein} and we arrive at the following form:
    \begin{widetext}
    \begin{equation}
     \langle T_K  (F_0F_1^{-1})^\epsilon(t^\eta)(F_0F_2^{-1})^{\epsilon'}(t'^{\eta'})\rangle_0
    = e^{i\epsilon\epsilon'\frac{\theta}{2}\sgn(X_1-X_2) \chi_{\eta,\eta'}(t-t')} = e^{-i\epsilon\epsilon'\frac{\theta}{2}\chi_{\eta,\eta'}(t-t')}
    \end{equation}
    \end{widetext}
where we used the definition of $\chi_{\eta,\eta'}(t-t')$
introduced for the contour ordered propagator of chiral bosons in
Eq.~\eqref{eq:chi}.

\end{document}